\documentclass[fleqn,usenatbib]{mnras}
\usepackage{newtxtext,newtxmath}
\usepackage[T1]{fontenc}
\usepackage{ae,aecompl}
\usepackage{graphicx}
\usepackage{float}
\usepackage{gensymb}
\usepackage{subfig}
\usepackage{natbib}
\usepackage{url}
\usepackage{amsmath}
\usepackage[scientific-notation=true]{siunitx}
\usepackage{caption}
\usepackage{array}
\usepackage{longtable}
\usepackage[toc,page]{appendix}

\title[The Photometric Selection of M-dwarfs]{The Photometric Selection of M-dwarfs using {\em Gaia}, WISE and 2MASS photometry}

\author[J. Bentley et al.]{J. Bentley$^{1}$\thanks{E-mail: j.bentley@unsw.edu.au}, C. G. Tinney$^{1,2}$, S. Sharma$^{4}$ and D. Wright$^{1,2,3,5}$\\
$^{1}$Exoplanetary Science at UNSW, School of Physics, UNSW Sydney, NSW 2052, Australia\\
$^{2}$Australian Centre for Astrobiology, UNSW Sydney, NSW 2052, Australia\\
$^{3}$Australian Astronomical Observatory, 105 Delhi Road, North Ryde, NSW 2113\\
$^{4}$Sydney Institute for Astronomy, School of Physics, A28, The University of Sydney, NSW, 2006, Australia\\
$^{5}$University of Southern Queensland, Computational Engineering and Science Research Centre, Toowoomba, Queensland 4350, Australia}
\begin{document}

\date{Accepted XXX. Received YYY; in original form ZZZ}

\pubyear{2018}

\label{firstpage}
\pagerange{\pageref{firstpage}--\pageref{lastpage}}
\maketitle

\begin{abstract}
We present criteria for the photometric selection of M-dwarfs using all-sky photometry, with a view to identifying M-dwarf candidates for inclusion in the input catalogues of upcoming all-sky surveys, including TESS and {\em FunnelWeb}.
The criteria are based on {\em Gaia}, WISE and 2MASS all-sky photometry, and deliberately do not rely on astrometric information. In the lead-up to the availability of truly distance-limited samples following the release of {\em Gaia} DR2, this approach has the significant benefit of delivering a sample unbiased with regard to space velocity. Our criteria were developed by using {\em Galaxia} synthetic galaxy model predictions to evaluate both M-dwarf completeness and false-positive detections (i.e. non-M-dwarf contamination rates). We have derived two sets of {\em Gaia} G\,\textless\,14.5 criteria -- a ``high-completeness'' set that contains 78,340 stars, of which 30.7-44.4\% are expected to be M-dwarfs and contains 99.3\% of the total number of expected M-dwarfs; and a ``low-contamination'' set that prioritises the stars most likely to be M-dwarfs at a cost of a reduction in completeness. This subset contains 40,505 stars and is expected to be comprised of 58.7-64.1\% M-dwarfs, with a completeness of 98\%. Comparison of the high-completeness set with the TESS Input Catalogue has identified 234 stars not currently in that catalogue, which  preliminary analysis suggests could be useful M-dwarf targets for TESS. We also compared the criteria to selection via absolute magnitude and a combination of both methods. We found that colour selection in combination with an absolute magnitude limit provides the most effective way of selecting M-dwarfs en masse. 
\end{abstract}


\section{Introduction}
M-dwarfs comprise around 75\% of the stars in our Milky Way \citep{2007Tarter}, making them the most common type of star in our Galaxy. Up to 50\% of these stars may harbour terrestrial exoplanets \citep{2013Kopparapu}. The low masses and luminosities of M-dwarfs also mean that their habitable zones correspond to relatively short-period orbits (\textless\,100\,d), making them prime targets for upcoming exoplanet transit surveys, \citep[e.g. NASA's Transiting Exoplanet Survey Satellite (TESS);][]{2009Ricker}, and ground-based Doppler follow-up to measure planetary masses and densities.\\

TESS will download full-field images sampled every 30 minutes \citep{2009Ricker}, allowing it to detect potentially habitable M-dwarf exoplanets over the whole sky down to around 14th magnitude. In addition, it will download data for a constrained number of target stars ($\sim$\,200,000) in its high cadence mode (sampling every 2 minutes). Clearly, it is desirable to know as much as possible about the best, high priority M-dwarf candidates in advance of TESS starting its survey. In the absence of spectroscopic data for every star in the sky down to TESS's faint magnitude limit, the next best strategy is selecting likely M-dwarfs using either astrometric or photometric information.\\

The use of astrometric selection has dominated M-dwarf identification to date (e.g. \citealt{1979Luyten}; \citealt{2000Zacharias}), with proper motion (and especially the reduced proper motion H; \citealt{1922Luyten}) being used to identify stars likely to be close to the Sun (and therefore low in luminosity). This does have the disadvantage, however, of only probing two components of each star's three-dimensional space motion, and so necessarily produces a biased sample.\\

Photometric M-dwarf selection suffers from contamination by non-M-dwarfs due to photometric uncertainties, source contamination and the intrinsic scatter of stellar properties about any adopted relationships between observed colours and physical parameters. On the other hand, they do not suffer any kinematic bias, and there are now very large, high-quality and publicly available all-sky databases at the optical and near-infrared wavelengths best suited to M-dwarf selection. This paper, therefore, explores the photometric approach.\\

The standard practice for fully and completely identifying M-dwarfs is through initial candidate selection using either photometry or astrometry (or both), followed by spectral classification. To date, obtaining spectroscopy for every possible M-dwarf in the sky down to the relevant magnitudes for TESS (i.e. {\em Gaia} G$\approx$14.5, I$\approx$14) has been infeasible. No instrument has had the combination of a sufficiently wide-field (5-10$\degree$ diameter field-of-view), aperture in the $>$1\,m range, and multi-object spectroscopic capability for hundreds of objects across that wide-field.\\

This is now changing. Both the {\em FunnelWeb}\footnote{http://funnel-web.wikispaces.com} survey in the southern hemisphere, and LAMOST \citep{1996Chu} in the northern hemisphere, can observe hundreds of targets (or more) over fields of 30 square degrees or more, with reconfiguration times between fields measured in minutes. These dedicated survey facilities now make it possible to observe tens of thousands of stars per night (or millions of stars per year), making possible a new generation of magnitude-limited, all-sky survey.\\

{\em FunnelWeb} (Rains et al. 2018, in prep.) is is a multi-object stellar survey of the Southern Hemisphere set to commence observations in July 2018. It will cover the entire southern sky (excluding only the most crowded regions with $\delta \leq 0\degree$, $|b| \geq 10\degree$) and  obtain high-quality (S/N~100) optical spectra for some $\sim$\,1.8 million stars down to a magnitude of {\em Gaia} G=14.5, aiming for 99\% completeness at the G=12.5 level. The survey is enabled by the TAIPAN instrument on the recently refurbished 1.2$\,$m UK-Schmidt Telescope at Siding Spring Observatory, which is able to simultaneously robotically position 150 optical fibres (or ``Starbugs'') within a 6\degree field of view. The instrument operates over the wavelength range 3700-8700$\hbox{\AA}$, and has a spectral resolution of R $\geq$ 2000. The main goals of the survey include a spectral library with detailed stellar parameters (including T$_{\rm eff}$, log(g), [Fe/H] and [$\alpha$/Fe]), including stellar spectra for targets being observed by the TESS satellite, with M-dwarfs a particular focus.\\ 

The combination of these new facilities, and the availability of large, deep and all-sky multi-colour surveys like {\em Gaia} \citep{2016Gaia}, the Two Micron All-Sky Survey (2MASS; \citealt{2006Skrutskie}), the Wide-field Infrared Survey Explorer data (WISE; \citealt{2010Wright}) and the Skymapper survey \citep{2007Keller} makes it eminently plausible to use of photometric M-dwarf selection, followed by the spectroscopic confirmation, for extremely large numbers of candidate M-dwarfs. However, even with the ability to observe a million stars per year, it is clearly still desirable to refine the criteria for selection of M-dwarf candidates as much as possible -- and in particular to understand the completeness and contamination that can be expected from a given set of photometric selection criteria. This is the focus of this paper.\\

\section{Data}
\subsection{Survey Data}
\label{Data}
\begin{figure}
	\centering
	\includegraphics[width=0.5\textwidth]{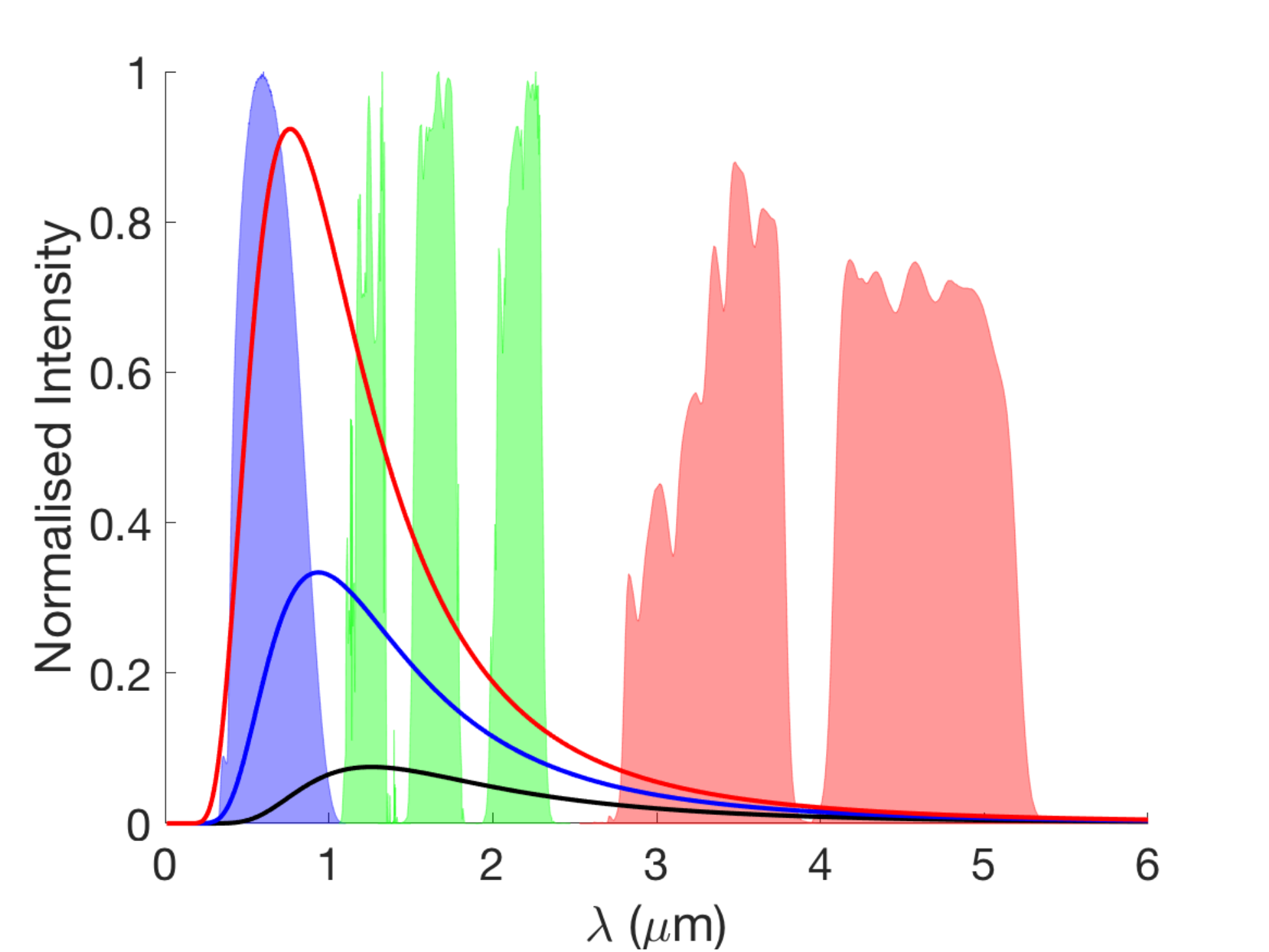}
 	\caption{The wavelength coverage of the {\em Gaia} G (blue), 2MASS J, H and K bands (green) and the WISE W1 and W2 bands (red) with black body curves at temperatures 3800, 3100 and 2300K that are each representative of a M0, M4 and M9 M-dwarf continuum (red, blue and black lines respectively).}
	\label{figband}
\end{figure}

The {\em Gaia}, WISE and 2MASS all-sky surveys provide photometry at a range of optical and near-infrared wavelengths. In particular, as can be seen from Figure\,\ref{figband}, they probe critical wavelengths for the selection of M-dwarfs, which emit most of their flux between 0.6\,nm and 5\,$\mu$m). \\

We selected a working dataset of all objects from {\em Gaia} DR1 \citep{2016GaiaDR1} with G\,$<$\,14.5 and cross-matched against the 2MASS PSC and {\em WISE} AllWISE catalogues using the {\em Gaia} DR1 portal ADQL\footnote{http://gea.esac.esa.int/archive/}. We limited the area of sky selected to that accessible from the southern hemisphere ($\delta$\,\textless\,+10\,\degree) and avoided the Galactic plane ($|$b$|>$10\,\degree) -- the latter criterion was adopted because the large WISE full-width-at-half-maximum of $\approx$\,6\," makes it difficult to match, or rely on, its data in crowded fields. Any object flagged in the AllWISE catalogue as a galaxy (xscproc\,$\neq$\,null), an extended object (ext\_flg\,$\neq$\,0), as multiple objects (n\_2mass\,\textgreater\,1) or as objects with poor/contaminated photometry (cc\_flags\,$\neq$\,0000) was also removed. The resulting catalogue contains almost 9 million stars, with 6.2 million stars at G\,\textless\,14.5, and Figure \ref{figScatterA} shows the number density of resulting sources in an example J--W1/W1--W2 colour-colour plane.\\

\subsection{Spectroscopic Comparison Sample}
\label{SpecData}
To provide a comparison set of photometry for stars with known spectral classifications, we extracted the same Gaia, 2MASS and WISE photometry for the large sample of M-dwarfs spectrally classified by \cite{2011West}, supplemented by the late K-dwarfs classified by \cite{2015Zhong}. To ensure the most precise photometry was used for these known M- and K-dwarfs, we only included stars in this comparison sample if they were quiescent for the bands we used using (var\_flg = XX\_\_, where X is from 0-5). Otherwise this comparison sample was selected in the same manner as our main sample.\\

\begin{figure*}
	\centering
    \subfloat[]{\label{figRelGJ}\includegraphics[width=0.39\textwidth]{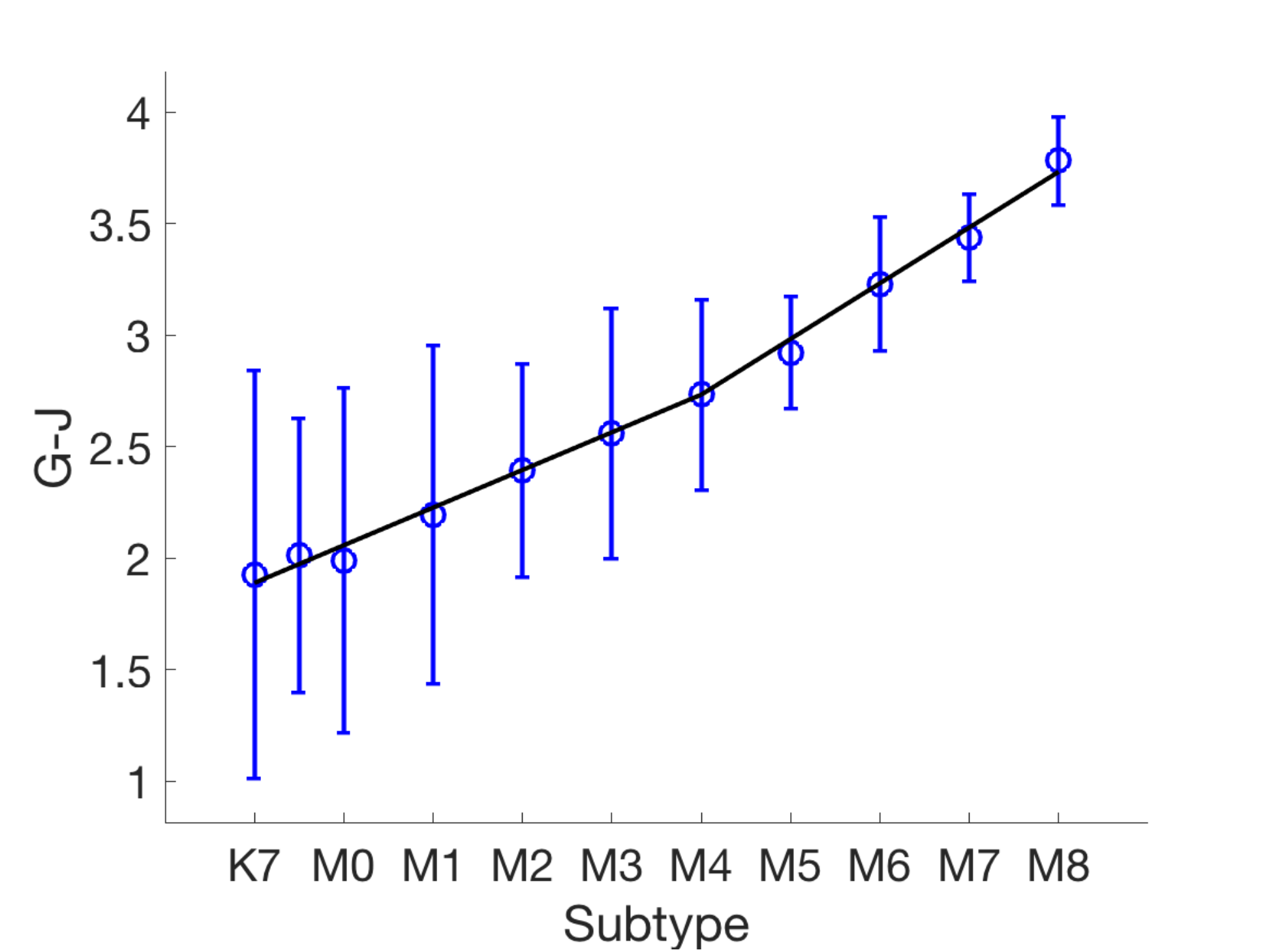}}
    \subfloat[]{\label{figRelGK}\includegraphics[width=0.39\textwidth]{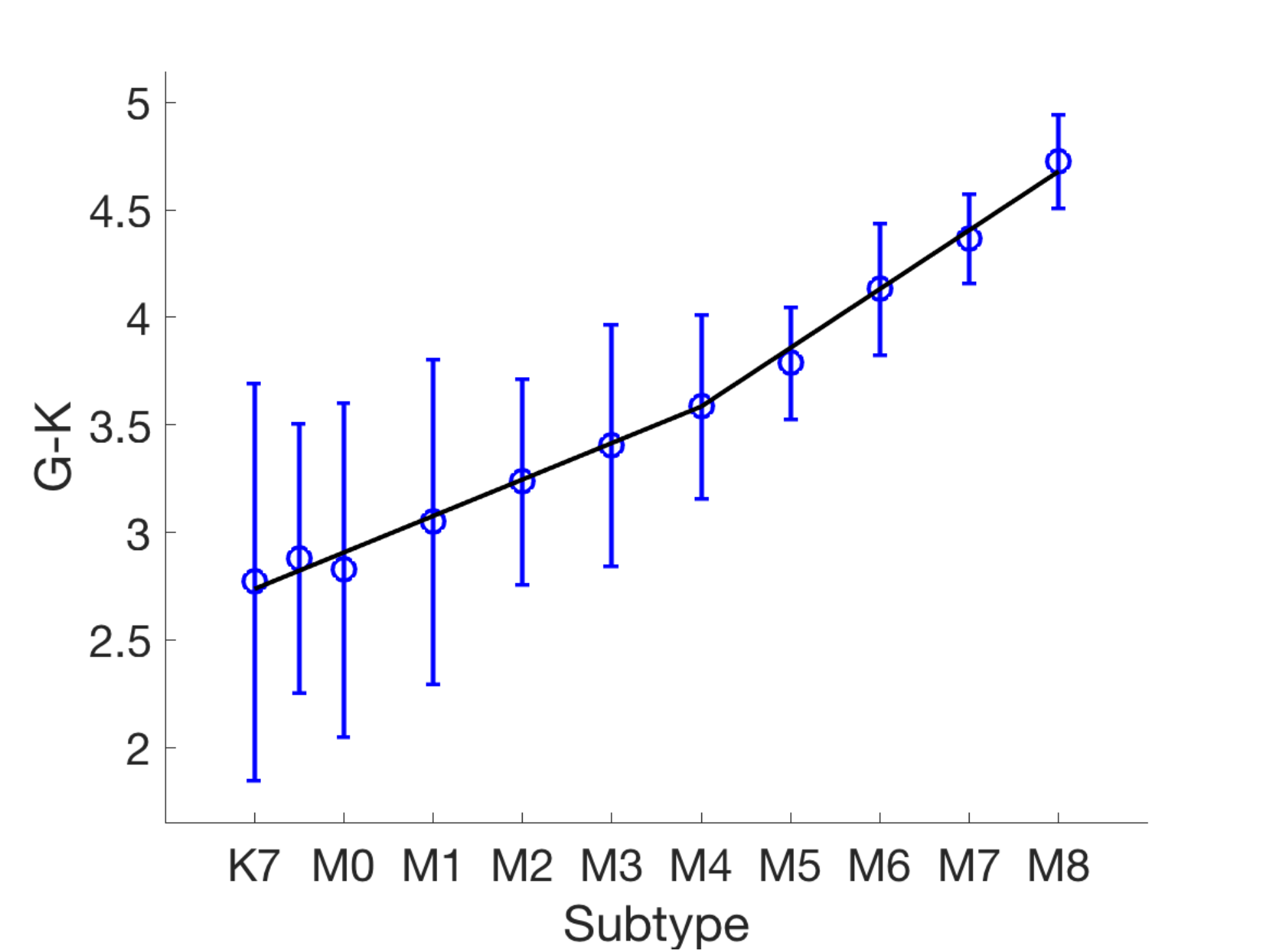}}\\
    \subfloat[]{\label{figRelKW2}\includegraphics[width=0.39\textwidth]{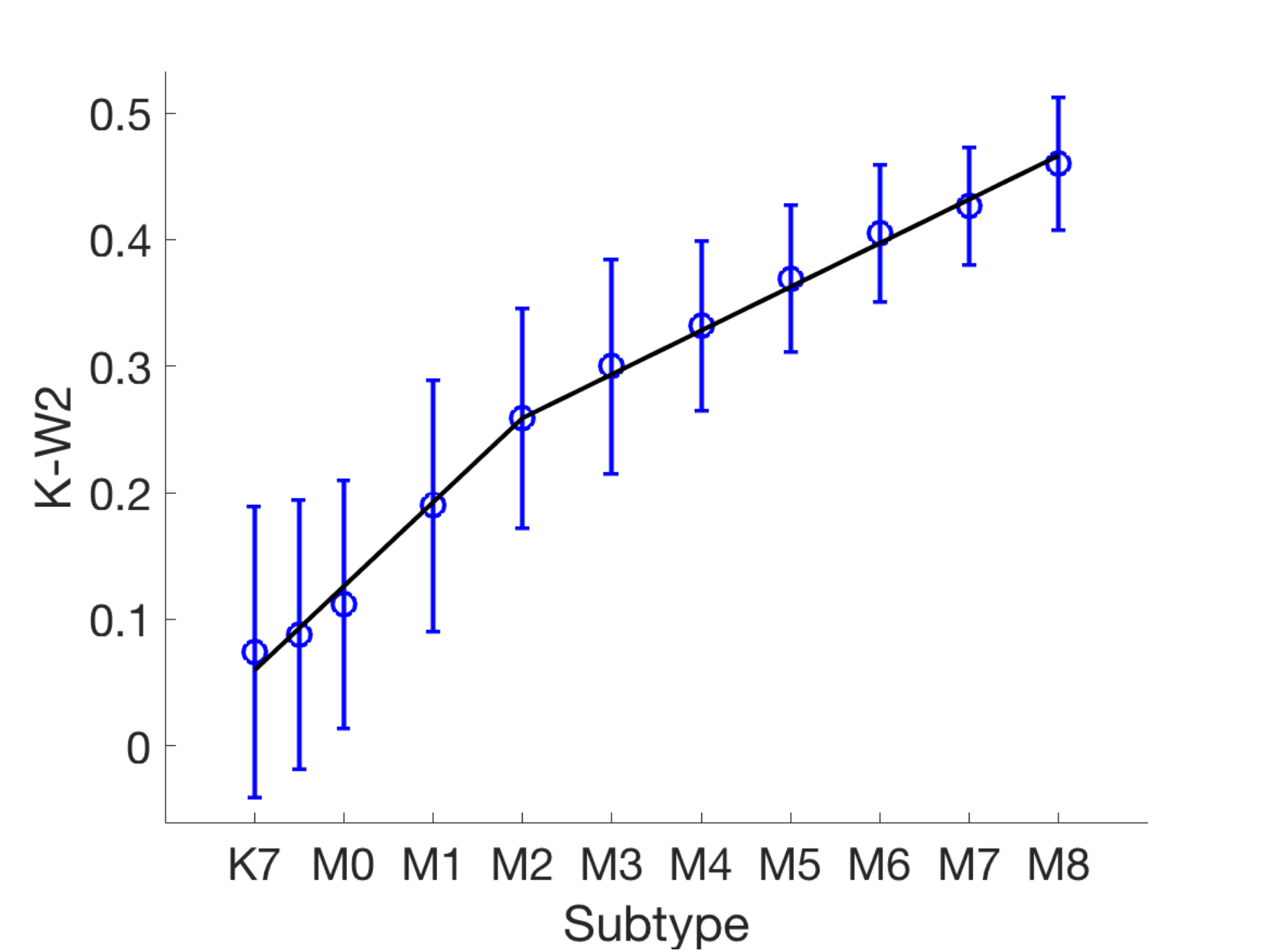}}
    \subfloat[]{\label{figRelW1W2}\includegraphics[width=0.39\textwidth]{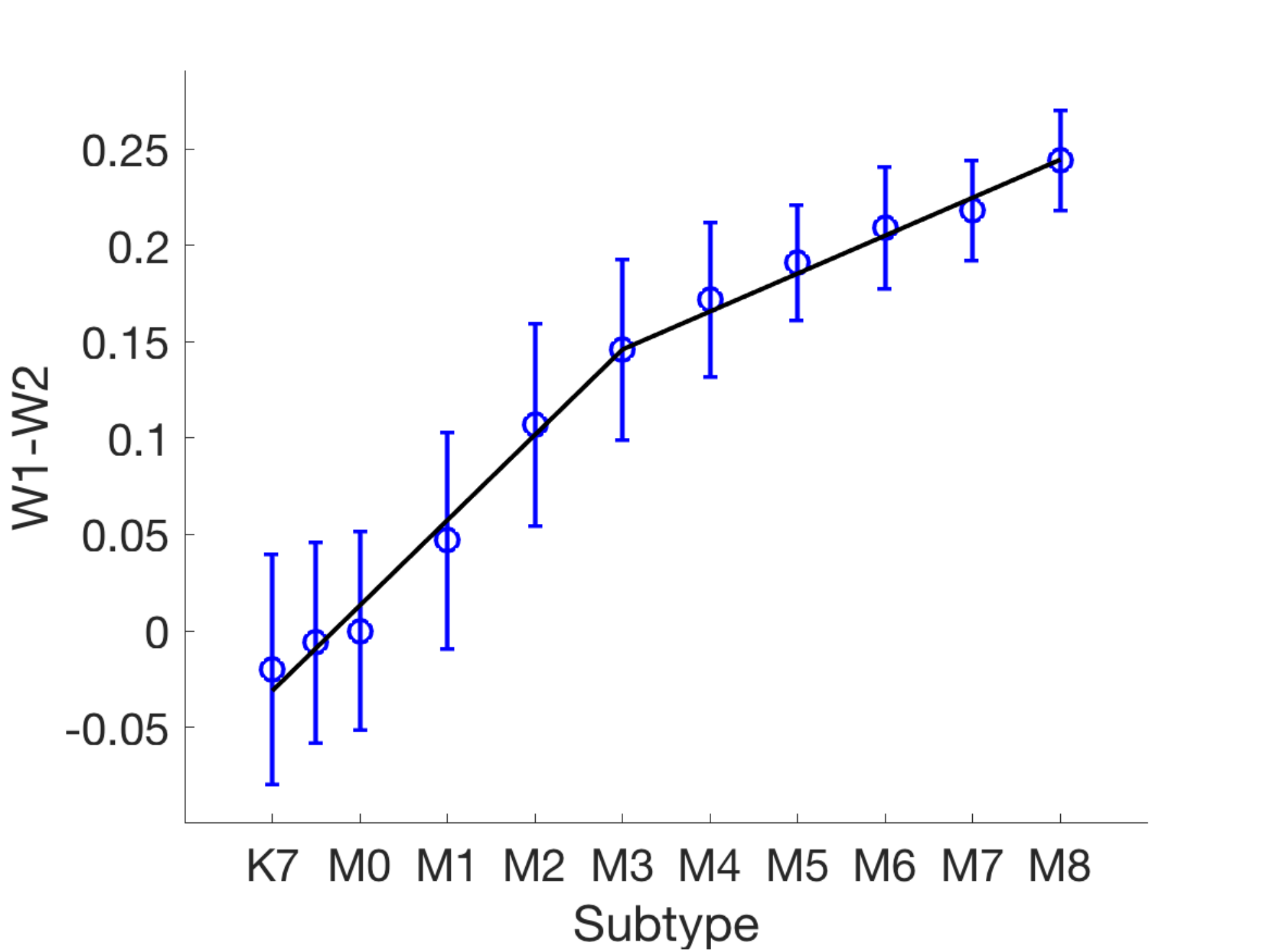}}\\
    \subfloat[]{\label{figRelJK}\includegraphics[width=0.39\textwidth]{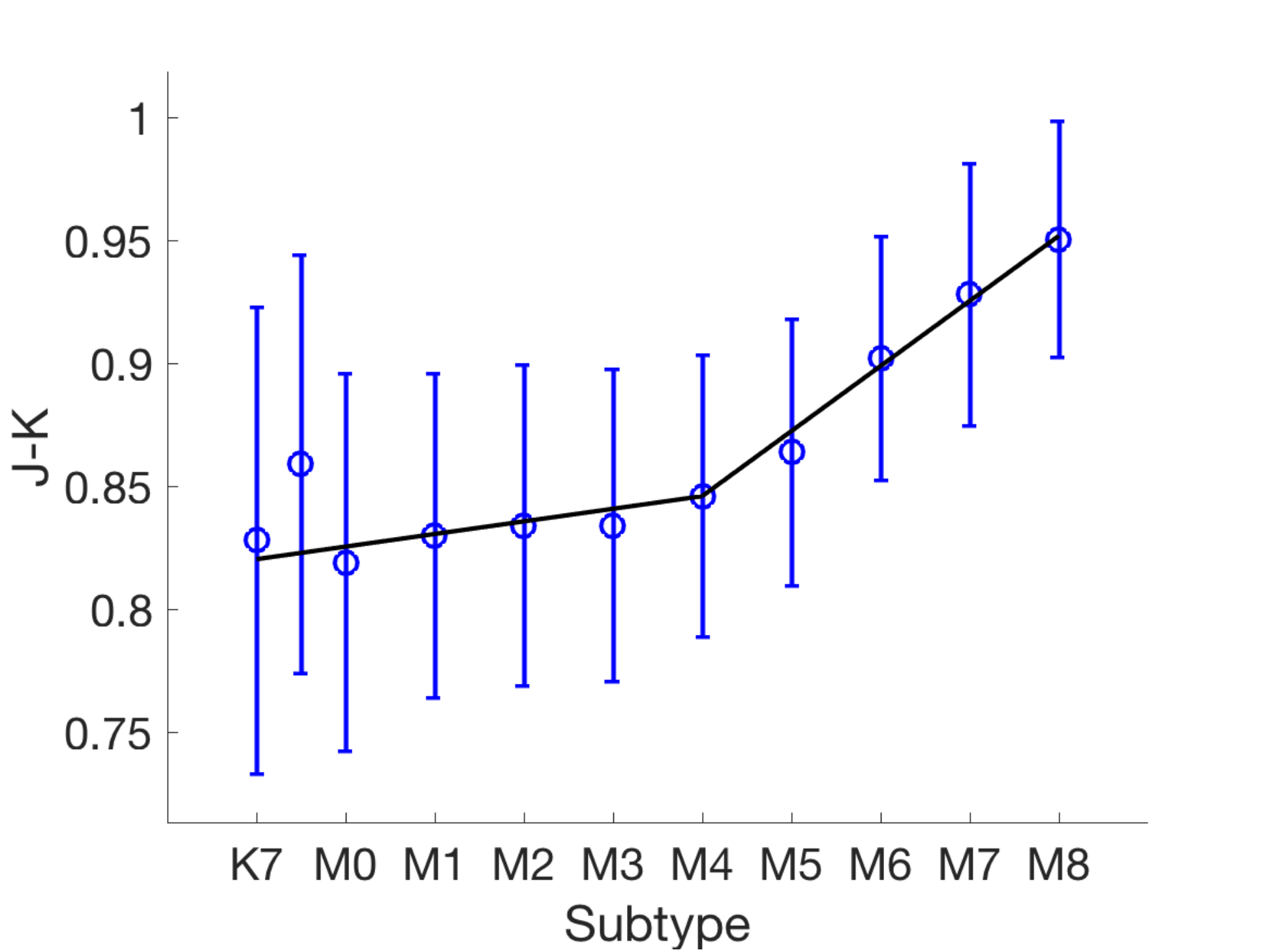}}
    \caption{Photometric Gaia/2MASS/WISE colours as a function of spectral type for M-dwarfs of West et al. (2011) and late K-dwarfs of Zhong et al. (2015). Circles are the median colour values for each subtype, with uncertainty bars showing r.m.s scatter. The solid black lines are linear fits to the median subtypes. The outlier K7.5 median colour value for J--K was excluded when determining that parameteristion.}
	\label{FigRelationship}
\end{figure*}

We calculated median colours and the r.m.s. scatter about the mean for each spectral type and these are plotted for J--W1/W1--W2 in Figure\,\ref{figSubtypes}, along with linear parametrisations of the median colours. The results of these parametrisations are tabulated in Table\,\ref{TabLookup} as a function of spectral type and seen in Figure\,\ref{FigRelationship}. In general the photometric scatter around these linear relationships are large -- usually several tenths of a magnitude -- with the scatter for optical-infrared colours becoming as large as seven-tenths of a magnitude at early types. This is much larger than expected due to the photometric measurement uncertainty of these surveys, suggesting this scatter is dominated by cosmic scatter due to metallicity variations, age variations, unresolved binarity, etc. Nonetheless, the trends with spectral type are consistent and smoothly varying.

\begin{table*}
	\begin{tabular}{| l | c | c  c | c c | c c | c c | l |}
	\hline
	Type & N  & G--J & r.m.s. & G--K & r.m.s. & K--W2 & r.m.s. & W1--W2 & r.m.s. & Src\\
	\hline
	K7 & 421 & 1.89 & 0.808 & 2.737 & 0.81 & 0.06 & 0.11 & -0.031 & 0.059 & ZJ \\
	K7.5 & 311 & 1.974 & 0.771 & 2.822 & 0.773 & 0.093 & 0.107 & -0.009 & 0.057 & ZJ \\
	M0 & 1268 & 2.058 & 0.733 & 2.907 & 0.736 & 0.126 & 0.103 & 0.013 & 0.055 & WA \\
	M1 & 1121 & 2.227 & 0.658 & 3.077 & 0.662 & 0.193 & 0.095 & 0.057 & 0.051 & WA \\
	M2 & 1493 & 2.396 & 0.583 & 3.246 & 0.588 & 0.259 & 0.088 & 0.102 & 0.047 & WA \\
	M3 & 2012 & 2.564 & 0.508 & 3.416 & 0.514 & 0.294 & 0.08 & 0.146 & 0.043 & WA \\
	M4 & 1974 & 2.733 & 0.433 & 3.585 & 0.44 & 0.328 & 0.072 & 0.166 & 0.04 & WA \\
	M5 & 732 & 2.983 & 0.358 & 3.859 & 0.366 & 0.363 & 0.065 & 0.185 & 0.036 & WA \\
	M6 & 311 & 3.232 & 0.283 & 4.132 & 0.292 & 0.397 & 0.057 & 0.205 & 0.032 & WA \\
	M7 & 128 & 3.482 & 0.208 & 4.406 & 0.218 & 0.432 & 0.049 & 0.225 & 0.028 & WA \\
	M8 & 14 & 3.732 & 0.133 & 4.679 & 0.144 & 0.466 & 0.042 & 0.244 & 0.024 & WA \\
	\hline
	\end{tabular}
    \caption{Colour sequences (and r.m.s. scatter about them) for late-K- and M-dwarfs. Spectral type sources are: ZJ, Zhong et al. (2015); WA, West et al. (2011).}
    \label{TabLookup}
\end{table*}

\section{Simulations}
\label{Modelling}
To test colour-based selection criteria,  we created simulated  photometry of a synthetic population of stars using {\em Galaxia} \citep{2011Sharma}. This is a Galactic simulation code that generates a synthetic stellar population, including parameters for every star simulated such as their masses, ages, temperatures and simulated photometry. {\em Galaxia} generates its synthetic populations based on models for the Galaxy's stellar populations and its star formation history. It first generates a population with a set of basic physical parameters (position, distance, mass, age, metallicity) very similar to the Besan\c{c}on model \citep{2003Robin}  to generate the thin disc, the thick disc, the bulge and the halo populations (respectively). It then derives the resulting luminosity, effective temperature, photometric magnitudes and colours for each star, using the PARSEC-v1.2S isochrones \citep{2012Bressan, 2014Tang, 2014Chen, 2015Chen}, the NBC version of bolometric corrections  \citep{2014Chen}, and assuming Reimers mass loss with efficiency $\eta=0.2$ for RGB stars. This photometry then has corrections applied to simulate the effects of extinction \citep{2011Sharma}. These simulations therefore include the effects of cosmic scatter due to metallicity variation and extinction, but do not include photometric measurement scatter, or the effects of source confusion. The simulation generated for this work was initially created for a magnitude limit of G\,\textless\,16, so that it would remain complete to G\,=\,14.5 when subject to the impacts of extinction and photometric scatter.\\

Because the {\em Galaxia} population is generated based on physical parameters, we require relationships between those parameters (i.e. mass, radius, gravity, age, temperature) and predicted spectral types of interest. In particular, what gravities and effective temperatures correspond to M-dwarfs? We have adopted the temperature and gravity estimates of \citet[Table 4.1]{2005Reid} in this work -- specifically we call a {\em Galaxia} object an M-dwarf if it has 2250\,K\,\textless\,T\,\textless\,3900\,K and 4.2\,\textless\,$\log g$\,\textless\,5.4. We also apply a lower age limit of 500 million years (i.e. we do not attempt to determine whether any object younger than 500\,Myr is an M-dwarf or not). This identifies $\approx$\,20,000 stars as M-dwarfs within our G\,\textless\,14.5 {\em Galaxia} sample of $\approx$\,4.9 million stars. \\

\begin{figure}
	\centering
    \subfloat[{\em Gaia}/WISE/2MASS data.]{\label{figScatterA}\includegraphics[width=0.33\textwidth]{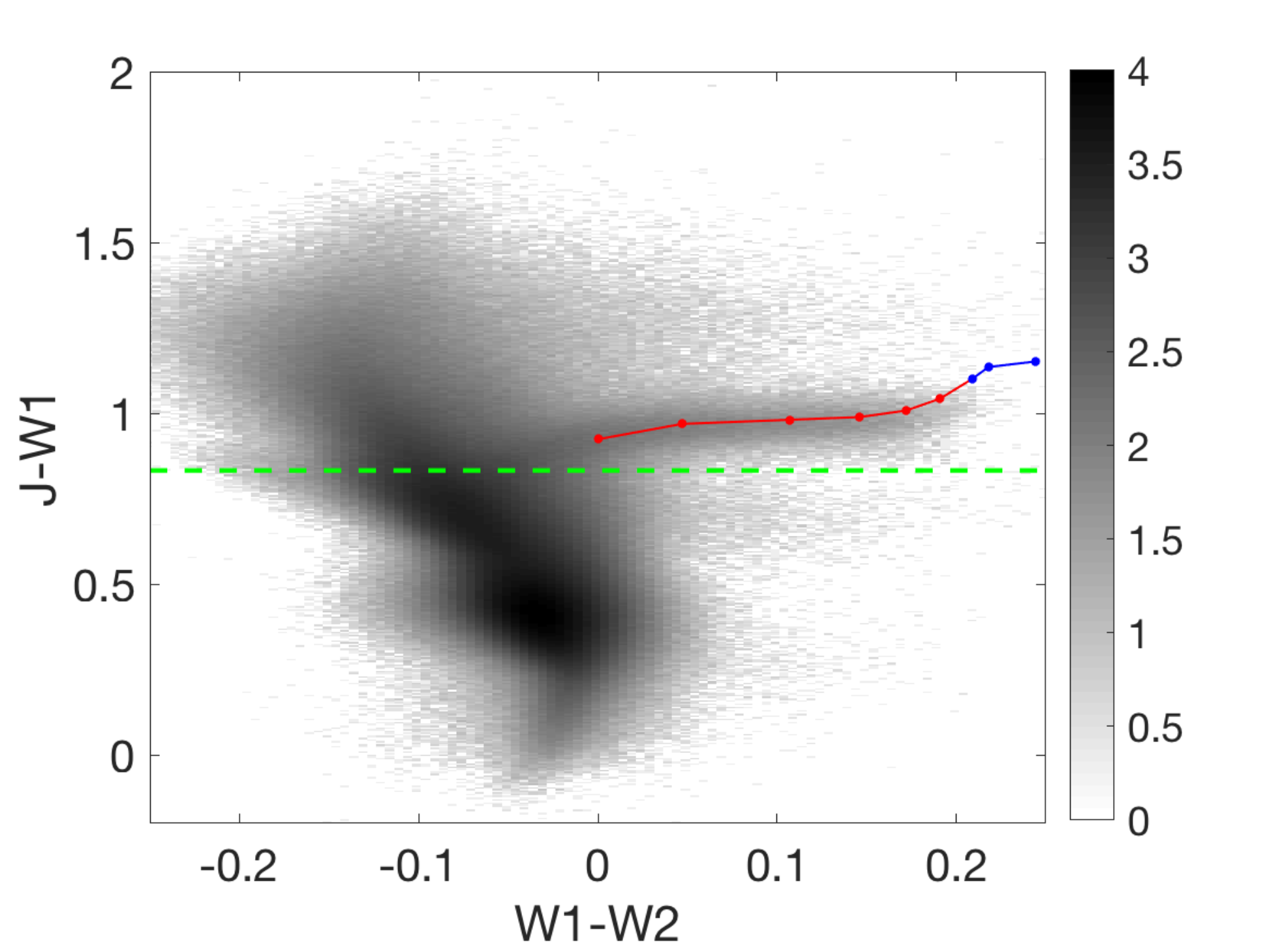}}\\[-1mm]
    \subfloat[{\em Galaxia} synthetic star population including cosmic scatter and extinction, but without photometric measurement scattering. M-dwarfs (see text) highlighted in red.]{\label{figScatterB}\includegraphics[width=0.33\textwidth]{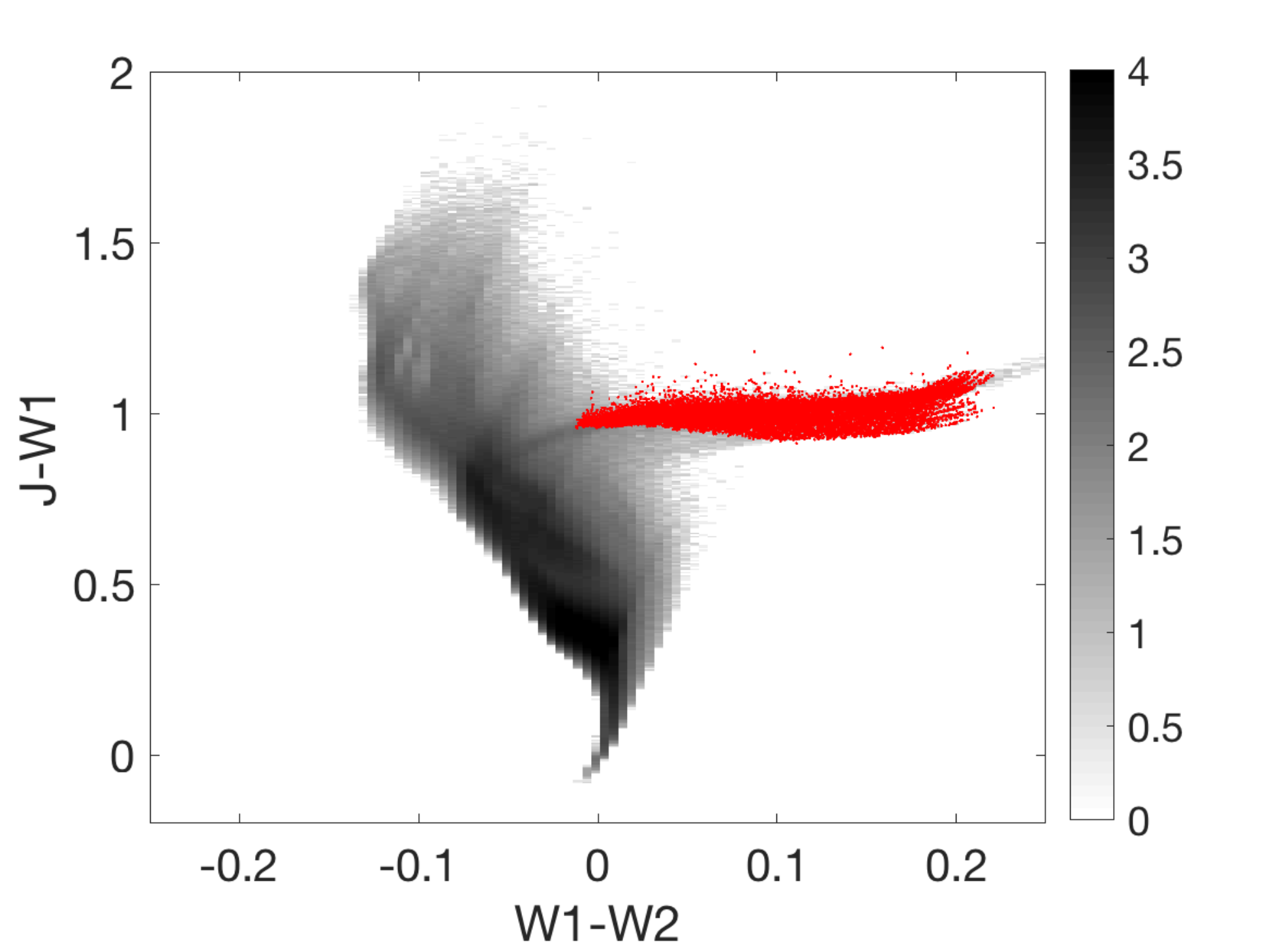}}\\[-1mm]
    \subfloat[{\em Galaxia} synthetic star population including cosmic scatter, extinction, and initial photometric measurement scattering.]{\label{figScatterC}\includegraphics[width=0.33\textwidth]{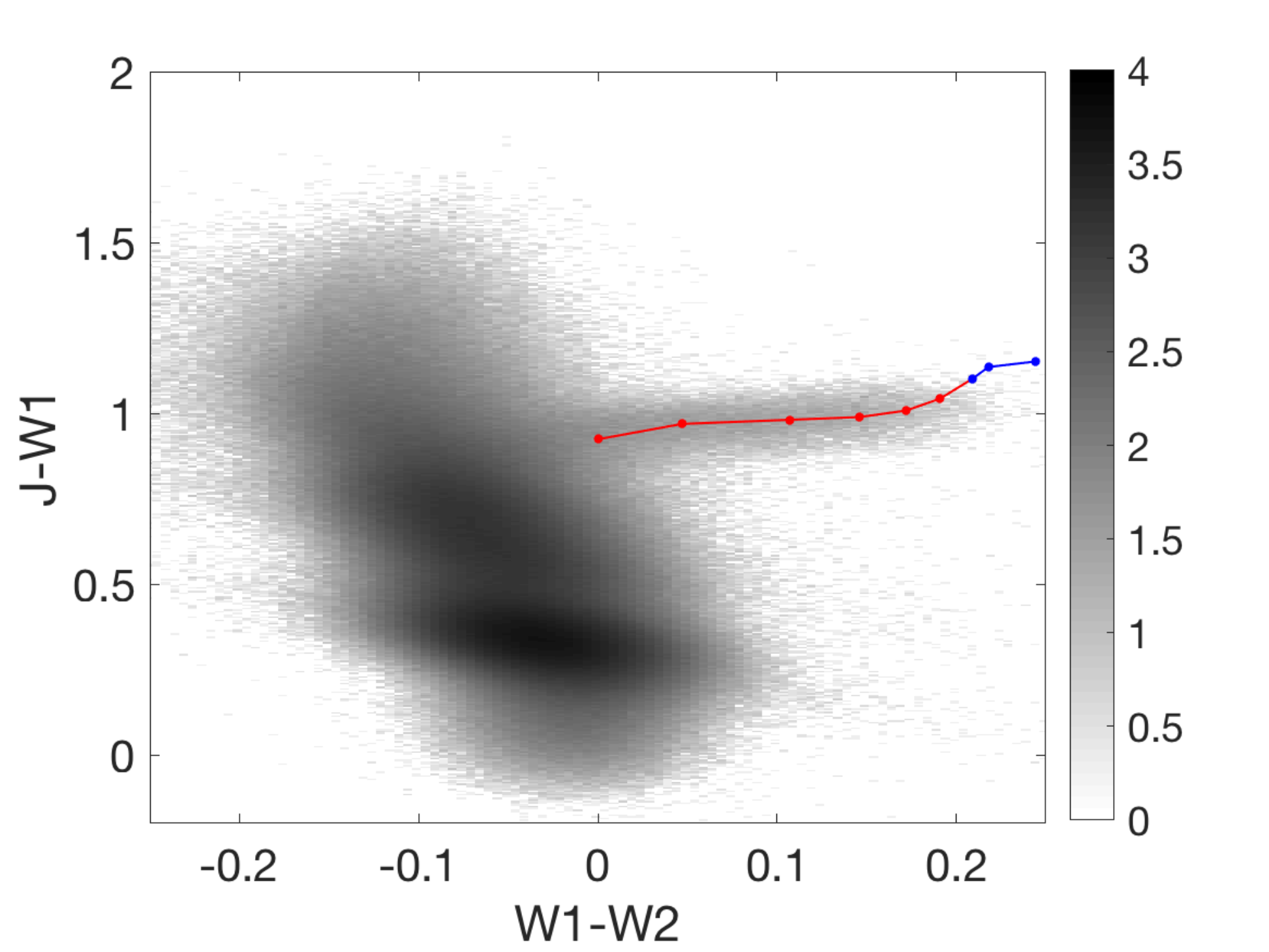}}\\[-1mm]
    \subfloat[{\em Galaxia} synthetic star population including cosmic scatter, extinction, and final photometric measurement scattering.]{\label{figScatterD}\includegraphics[width=0.33\textwidth]{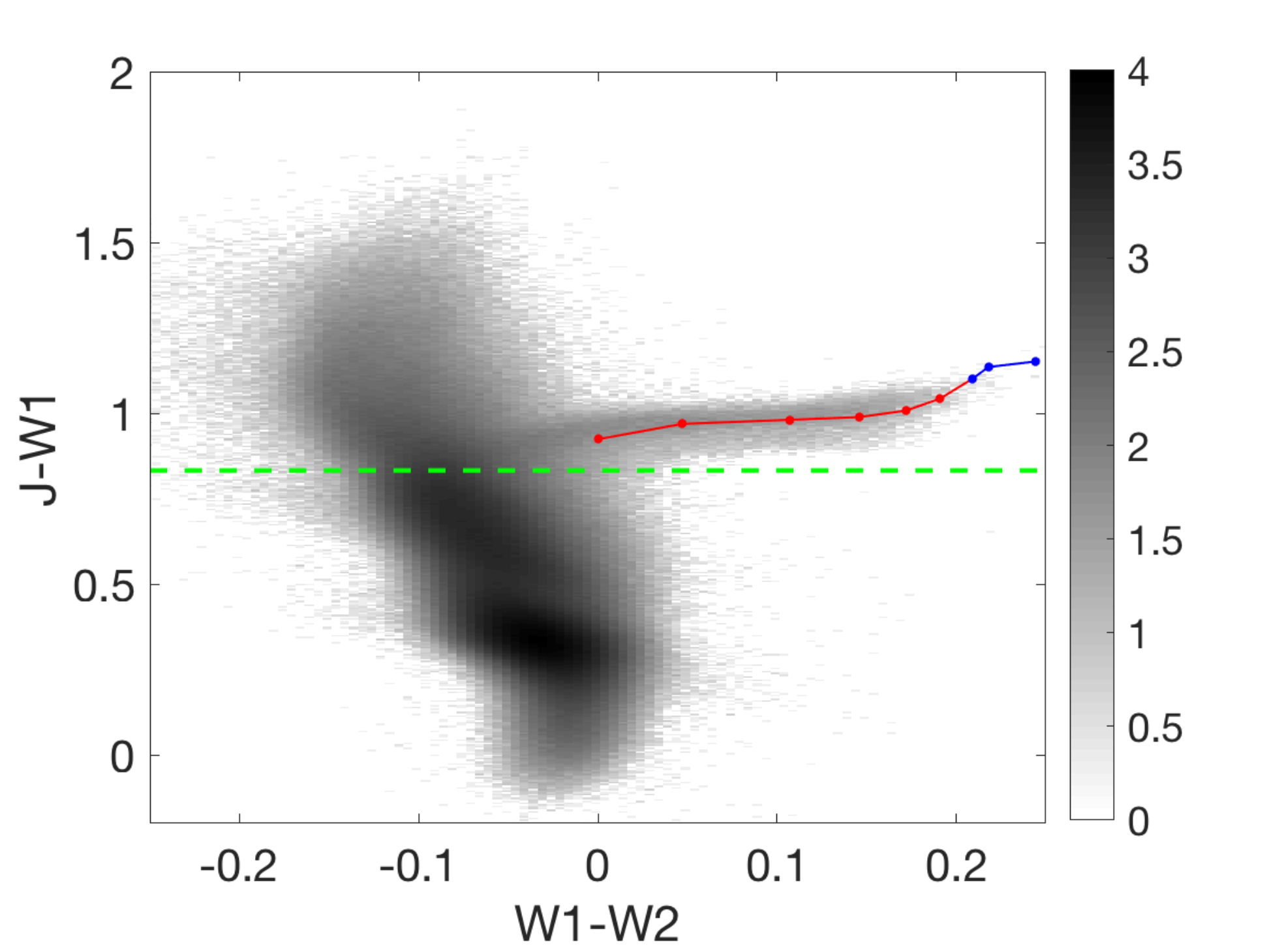}}
    \caption{Logarithmic source densities in the J--W1/W1--W2 plane for sources with 9\,\textless\,G\,\textless\,14.5. Objects below the green horizontal line in the \ref{figScatterA} and \ref{figScatterD} are used to normalise  {\em Galaxia} to observational data. The M-dwarf spectroscopic comparison sample from \S\,\ref{SpecData} are overplotted in red and blue.} 
    \label{figScatter}
\end{figure}

Figure\,\ref{figScatterB} shows a density plot of this {\em Galaxia} sample in the J--W1/W1-W2 plane, along with the objects identified as M-dwarfs by our adopted criterion (over-plotted in red).\\

\begin{figure}
	\centering
	\includegraphics[width=0.4\textwidth]{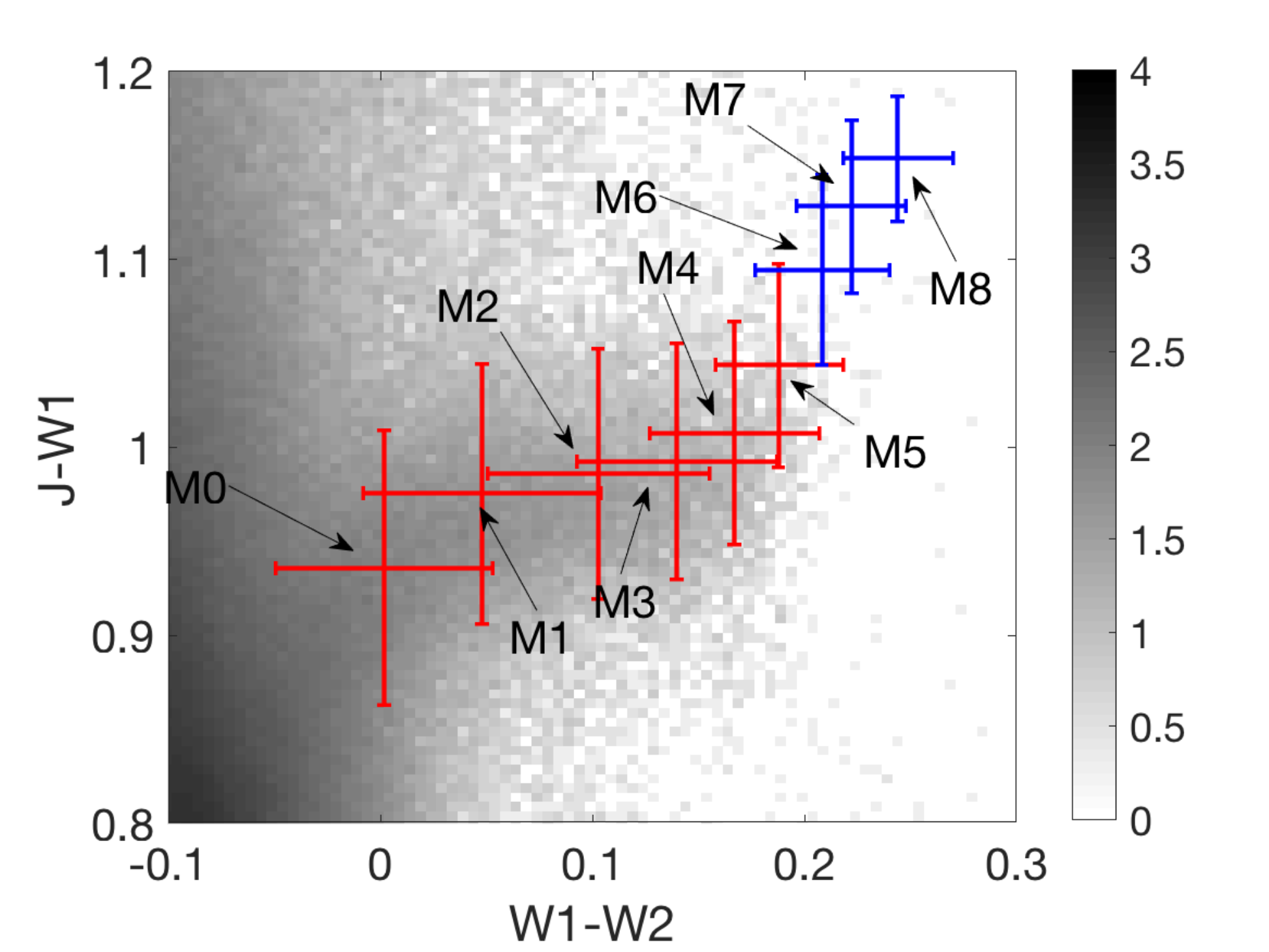}
    \caption{{\em Galaxia} generated colour-colour density in J--W1/W1--W2 expanded around the M-dwarf branch. Overplotted are points representing the median colour and r.m.s. for spectroscopically observed M-dwarfs (M0-M5 in red, M5-M8 in blue).}
    \label{figSubtypes}
\end{figure}

{\em Galaxia} is necessarily limited in its predictive power for the photometry of sources by its PARSEC stellar models. In particular, {\em Galaxia} cannot predict the photometric properties of the latest M-dwarfs, because they are not included in the PARSEC isochrones. Specifically there are no isochrones for dwarfs of later than M5 (i.e. for T\,\textless\,2700\,K with 4.2 $<$ $\log g$ $<$ 5.4). We highlight this by showing in Figure\,\ref{figSubtypes} an expanded region of Figure \ref{figScatterB} around the M-dwarf branch, along with the M-dwarf sequence for spectroscopically observed M-dwarfs  from \S\ref{SpecData}.\\

The latest spectroscopically observed M-dwarfs (M6-8 plotted in blue) lie in a region where {\em Galaxia} simulates almost no objects, because it {\em can} simulate no objects. \\

However, it should be noted that (1) the number of M dwarfs will drop dramatically at later spectral types in any magnitude limited sample, (2) the TESS survey (at least) will focus on M-dwarfs earlier than M5 \citep{2015Ricker},  and (3) these late M-dwarfs have photometric properties that are well-known from other work \citep{2002Leggett} and \S\ref{SpecData} making them quite easy to distinguish from M-giants and other types of stars\,\citep{2011Lepine}. As a result these ``missing'' {\em Galaxia} late M-dwarfs will make an insignificant contribution to our estimates of completeness and contamination.\\

\subsection{Photometric scatter}
\label{Scattering}
As noted above, scatter needs to be added to  {\em Galaxia}'s simulated photometry to match the expected photometric uncertainties for our observational data -- this is obvious from Figs. \ref{figScatterA} and \ref{figScatterB}, where we see the observed distribution is noticeably more scattered from the {\em Galaxia} one.\\

\begin{figure}
	\centering
    \includegraphics[width=0.4\textwidth]{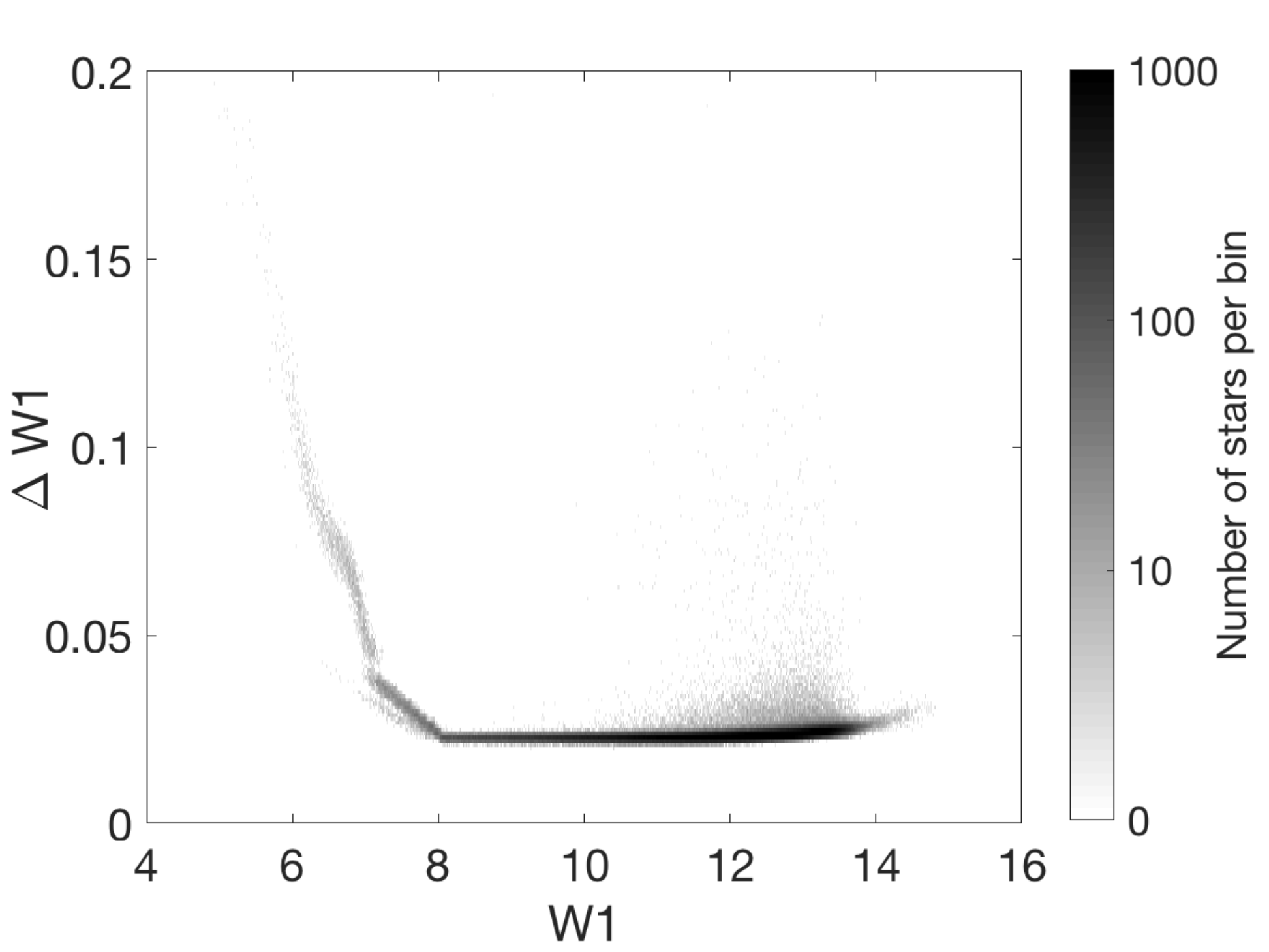}
    \caption{Density plot of the number of stars per bin of the WISE photometric uncertainty as a function of the WISE magnitude W1.}
    \label{figDelta}
\end{figure}

The observational data all come with estimated 1-$\sigma$ uncertainties, so we obtained uncertainty distribution functions for each observed bandpass by binning these uncertainties in 0.005 magnitude bins (for 2MASS/WISE) and  0.0002 magnitude bins (for  {\em Gaia}) -- see Figure\,\ref{figDelta} for an example. \\

For each simulated magnitude (e.g. $W1$), we then randomly drew from the appropriate distribution function for that magnitude bin, to obtain an estimate of the uncertainty for that simulated object (e.g. $\sigma_{W1}$). We then randomly drew from a normal distribution with that uncertainty to obtain a scattering estimate (e.g. $\Delta_{W1}$), which was applied to the simulated magnitude to obtain a scattered magnitude. Following the application of this suitable level of photometric scatter, our simulation was then trimmed to only contain simulated sources with G\,\textless\,14.5.\\

The result of this photometric scattering applied to the data of Figure\,\ref{figScatterB} is shown in Figure\,\ref{figScatterC} -- the result is to significantly ``over scatter'' the simulated data compared to the observational data of Figure\,\ref{figScatterA}. 
All the features in W1--W2 in the scattered simulated data are too broad, as is the width of the M-dwarf branch in the J--W1 direction. Since the aim of our simulations is (in the end) to create simulated data that looks like observational data, we empirically adjusted the fraction of the predicted photometric scatter applied to the simulated data, so as to matched the observed distribution. It was found our initial scattering needed to be reduced by a factor of near a half, and this is shown in Figure\,\ref{figScatterD}.\\

\subsection{Colour terms and offsets}
\label{secOff}
Comparison of the simulated and scattered data with the observational data, showed small -- but notable -- colour differences between  {\em Galaxia} predicted colours and observed colours at the level of several hundredths of magnitude to a few tenths of a magnitude. This is not entirely surprising, inasmuch as the {\em Galaxia} predicted colours rely entirely on synthetic model atmospheres and model filter profiles. \\

\begin{figure}
	\centering
    \includegraphics[width=0.4\textwidth]{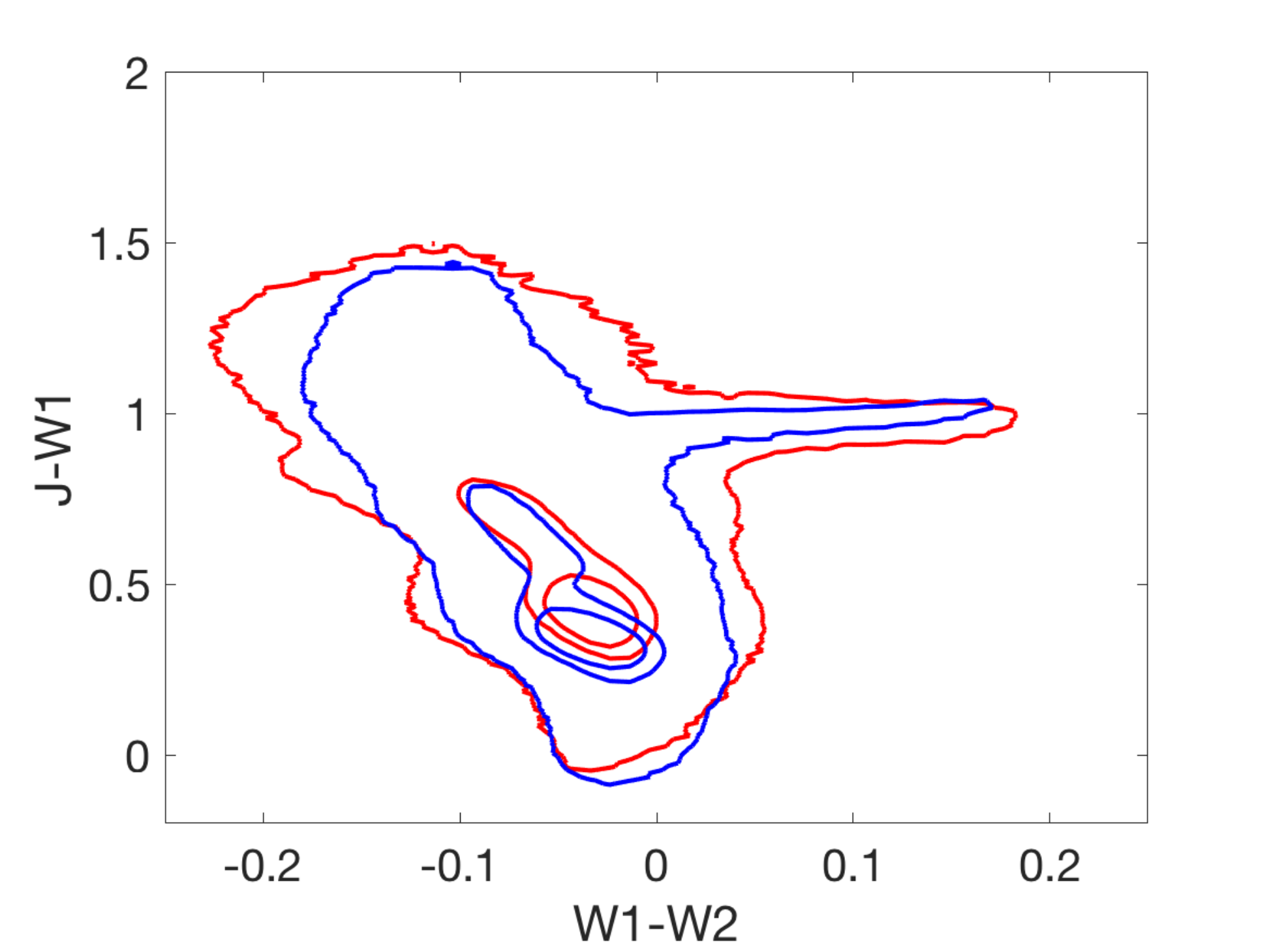}
    \caption{Contours containing 30\%, 60\% and 99\% of the observational (in red) and simulated (in blue) data in J--W1, W1--W2. Offsets from Table\,\ref{tabOffset} have been used to align the simulation M-dwarf branch to its observational counterpart.}
    \label{figLandmark}
\end{figure}

We therefore used ``benchmark'' features in the observed data, to align the two sets of data. Figure\,\ref{figLandmark} shows example contours (rather than grey scales) of source density in J--W1/W1--W2 after these offsets were determined and applied. Offsets were determined using observed stars with G=9-14.5 from a number of features: the M-dwarf branch at J--W1$\approx$1; the G-dwarf clump at W1--W2$\approx$-0.03, J--W1$\approx$0.35; and the giant clump at W1--W2$\approx$-0.07, J--W1$\approx$0.7. Because the aim of this work is to understand how selection criteria for M-dwarfs will be influenced by contamination, these offsets were weighted more heavily on the features nearest to the M-dwarf branch, as these are the source most likely to contaminate our M-dwarf selection. This is why in Figure\,\ref{figLandmark} the giant clump is better aligned than the G-dwarf clump. The offsets so determined and applied are reproduced in Table\,\ref{tabOffset}.\\

\begin{table}
	\begin{center}
	\begin{tabular}{c c c c c} 
 		\hline
 		 G--J & G--H & G--K & G--W1 & G--W2\\
 		-0.1 & -0.25 & -0.2  & -0.3  & -0.3\\
        \hline
            & J--H & J--K & J--W1 & J--W2\\
            & -0.01 & -0.01 & -0.03 & -0.07\\
            \cline{2-5}
            \cline{2-5}
            &    & H--K & H--W1 & H--W2\\
            &    & 0 & -0.02 & -0.05\\
            \cline{3-5}
            \cline{3-5}
            &    &    & K--W1 & K--W2\\
            &    &    & -0.04 & -0.02\\
            \cline{4-5}
            \cline{4-5}
            &    &    &     & W1--W2\\
            &    &    &     & -0.025\\
 		\cline{5-5}
	\end{tabular}
    \caption{Zero-point colour offsets to align {\em Galaxia} colours with observed colours (in the sense that these offsets are added to synthetic colours).}
    \label{tabOffset}
	\end{center}
\end{table}

\subsection{Normalising {\em Galaxia} to the Sky}
\label{SecNorm}
Each {\em Galaxia} simulation run produces an arbitrary number of stars (in this case 4,931,764). To compare this synthetic {\em Galaxia} population with the observational {\em Gaia}/WISE/2MASS data, we therefore need to estimate a normalisation between the two data sets. As noted above, while the colour offsets adopted (Table\,\ref{tabOffset}) aligned the M-dwarf branch, and features near that branch, in both datasets, there remain  a number of differences.\\

As noted above the G-dwarf clump does not align in Figure\,\ref{figLandmark}, even after the M-dwarf and giant clumps are aligned. We believe this to be due to limits on the reliability of the synthetic photometry adopted by {\em Galaxia}. Even after aligning for colour offsets there clearly remain higher-order colour terms.\\

Another difference between the observed and simulated data is seen in the M-giant region at J--W1$>$1 (see Figure\,\ref{figScatterA} and \ref{figScatterD}). The real galaxy produces a plume of objects with a wide distribution of colours, while {\em Galaxia} simulates a much narrower range. This is likely to be due to the significant variability and mass-loss present in this class of stars, resulting in intrinsic reddening and scattering in colour that is highly variable from source to source. In this region the simulated data also contains fewer stars than the observational data.\\

Given these differences it was felt that normalising {\em Galaxia} to the Galaxy using the total number of objects in both samples would not be the best way to proceed. Instead we compared the types of stars best represented in both datasets. {\em Galaxia} has been heavily tuned to match the observed Galaxy for relatively unreddened G-K dwarfs and giants. We therefore normalised the two samples using the total number of stars 0.1\,mag below the M-dwarf branch in each colour-colour space (shown by the green horizontal lines in Figures\,\ref{figScatterA} and \ref{figScatterD}), and adopted a normalisation factor of 1.211 to scale up the synthetic population to match the observational data. This means that, while the overall population will be close in number, specific regions may have differing numbers of stars. 

\section{Colour-colour selection}
We have followed the long-standing tradition in giant-dwarf discrimination of selecting colour-colour diagrams which aim to break degeneracies between stellar effective temperature and surface gravity. Specifically, we seek colours sensitive to effective temperature for the horizontal axes, and sensitive to surface gravity for the  vertical axes. Of course, no such diagram is perfect and contamination will occur where different classes of star overlap. Most notably for our case, the low-temperature boundary of the M-dwarf regime which adjoins the (much more numerous) high temperature boundary of the late K-dwarf regime, as well as the less numerous lower boundary of the K- and M-giants. Even small amounts of photometric scatter, cosmic scatter and source confusion in these more numerous populations will lead to significant contamination of any M-dwarf selection.\\

\subsection{Connected Component Analysis}
\label{secCCA}
\begin{figure}
	\centering
    \includegraphics[width=0.4\textwidth]{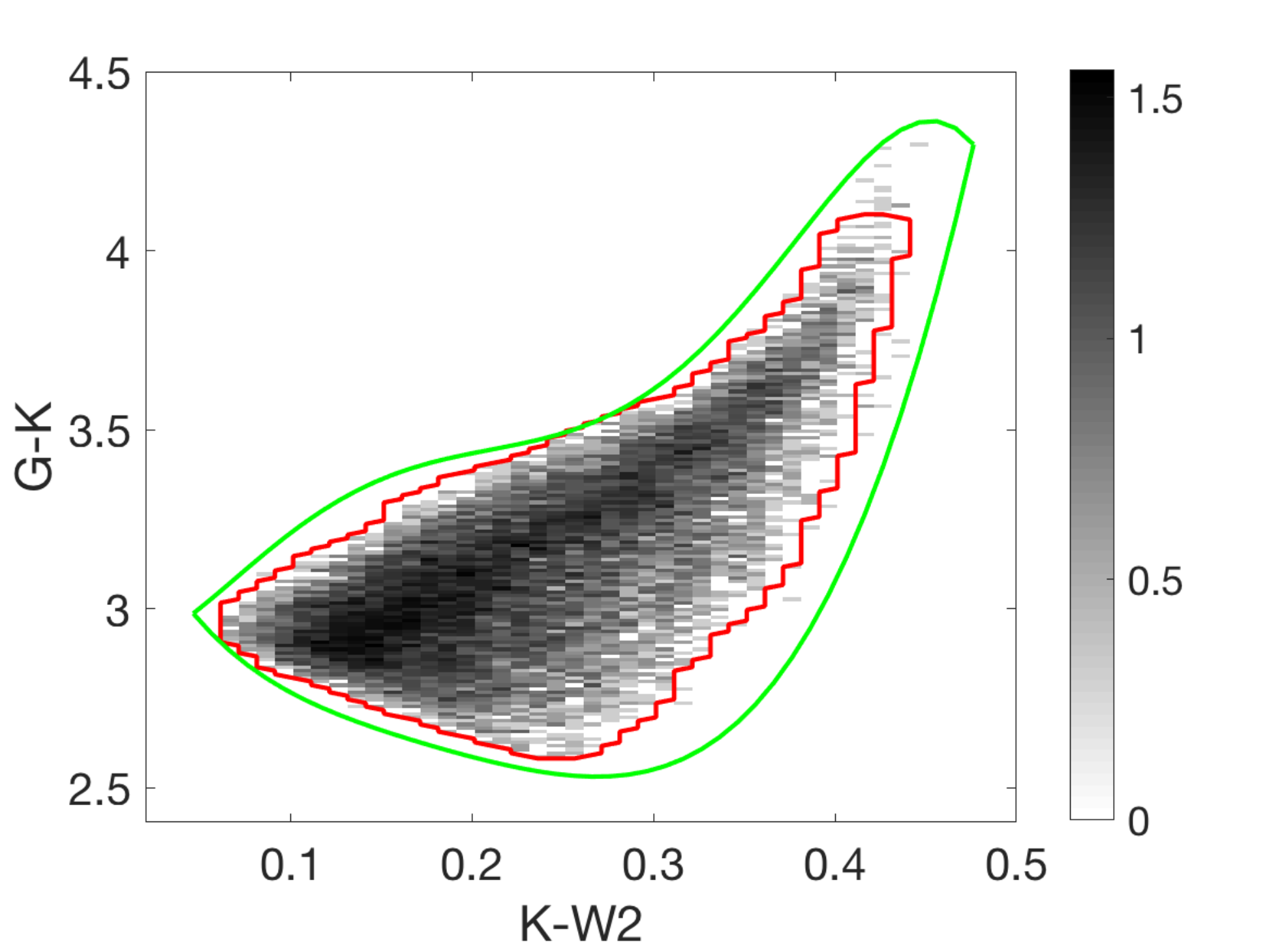}
    \caption{Connected component analysis of the {\em Galaxia} M-dwarfs using the G--K and K--W2 colours. The red outline highlights the region defined as the largest component while the green line represents the boundary region around the entire {\em Galaxia} M-dwarf population for this colour plane.}
    \label{figCCA}
\end{figure}

To aid in selecting colour-colour planes, we have made use of Connected Component Analysis (CCA; \citealt{1988Samet}) to identify the colour regions spanned by key populations. CCA is a technique that identifies the pixels of a 2-dimensional image that comprise each component of the image. For our case, the images of interest are density images created by binning {\em Galaxia} simulated objects in colour-colour planes. Using this technique, we can select groups of components that comprise the majority of our sample, excluding highly scattered components that comprise a small fraction of the total population but would inflate our selected region, adding significant numbers of stars we do not intend to select (i.e. non-M-dwarfs). For example, Figure\,\ref{figCCA} shows an example use of this technique for a density image of our adopted {\em Galaxia} M-dwarfs in the G--K/K--W2 colour-colour plane. The green contour is a rough measure of the region bound by all the {\em Galaxia} M-dwarfs while the red contour uses CCA to select 99.5\% of the M-dwarfs but excludes a number of the most photometrically scattered M-dwarfs. Losing 0.5\% of the M-dwarfs in our sample is an acceptable sacrifice to reduce the selected region and avoid as many non-M-dwarfs as we can.\\

We therefore compiled a set of criteria (listed in Table\,\ref{TabMK}) to identify {\em Galaxia} sources as M- or K-giants, or K-dwarfs (as well as the previously discussed criterion for M-dwarfs). We then used CCA to identify regions in our data that contain \textgreater\,97\% of each of these classes of object in our {\em Galaxia} simulations (see Figure\,\ref{figCCP}).\\

\begin{table}
\begin{center}
	\begin{tabular}{| c | c | c | c |}
		\hline
        Sp. Type & log(g) & Age\,(Yr) & T\,(K) \\
        \hline
        M-dwarf & 4.2\,-\,5.4 & \textgreater\,5\,x\,10$^{8}$ & 2250\,-\,3900 \\
        K-dwarf & 4.2\,-\,5.4 & \textgreater\,5\,x\,10$^{8}$ & 3900\,-\,4600 \\
        M-giant & \textless\,4.2 & \textgreater\,5\,x\,10$^{8}$ & 2250\,-\,3900 \\
        K-giant & \textless\,4.2 & \textgreater\,5\,x\,10$^{8}$ & 3900\,-\,4600 \\
        \hline
\end{tabular}
\caption{Stellar characteristics used to define the M and K, dwarf and giant populations in the synthetic {\em Galaxia} population.}
\label{TabMK}
\end{center}
\end{table}

\subsection{Stellar distributions}
After examining the many colour-colour combinations available to us from our six passbands, we selected three for use in our analysis. Each possesses distinct advantages. \\

\begin{figure}
	\centering
    \subfloat[]{\label{figColSpaceGKKW2}\includegraphics[width=0.4\textwidth]{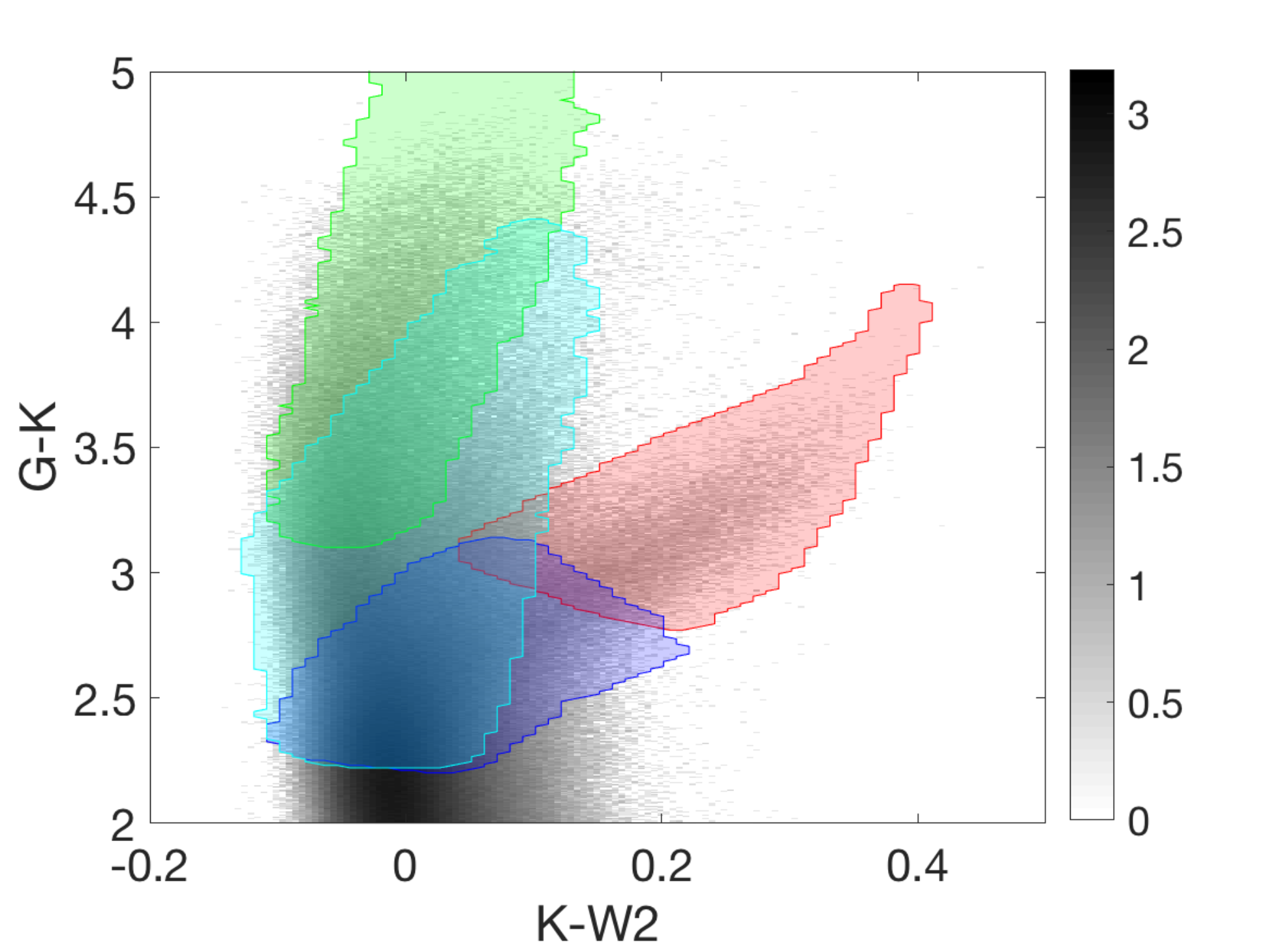}}\\[-1mm]
    \subfloat[]{\label{figColSpaceJKGJ}\includegraphics[width=0.4\textwidth]{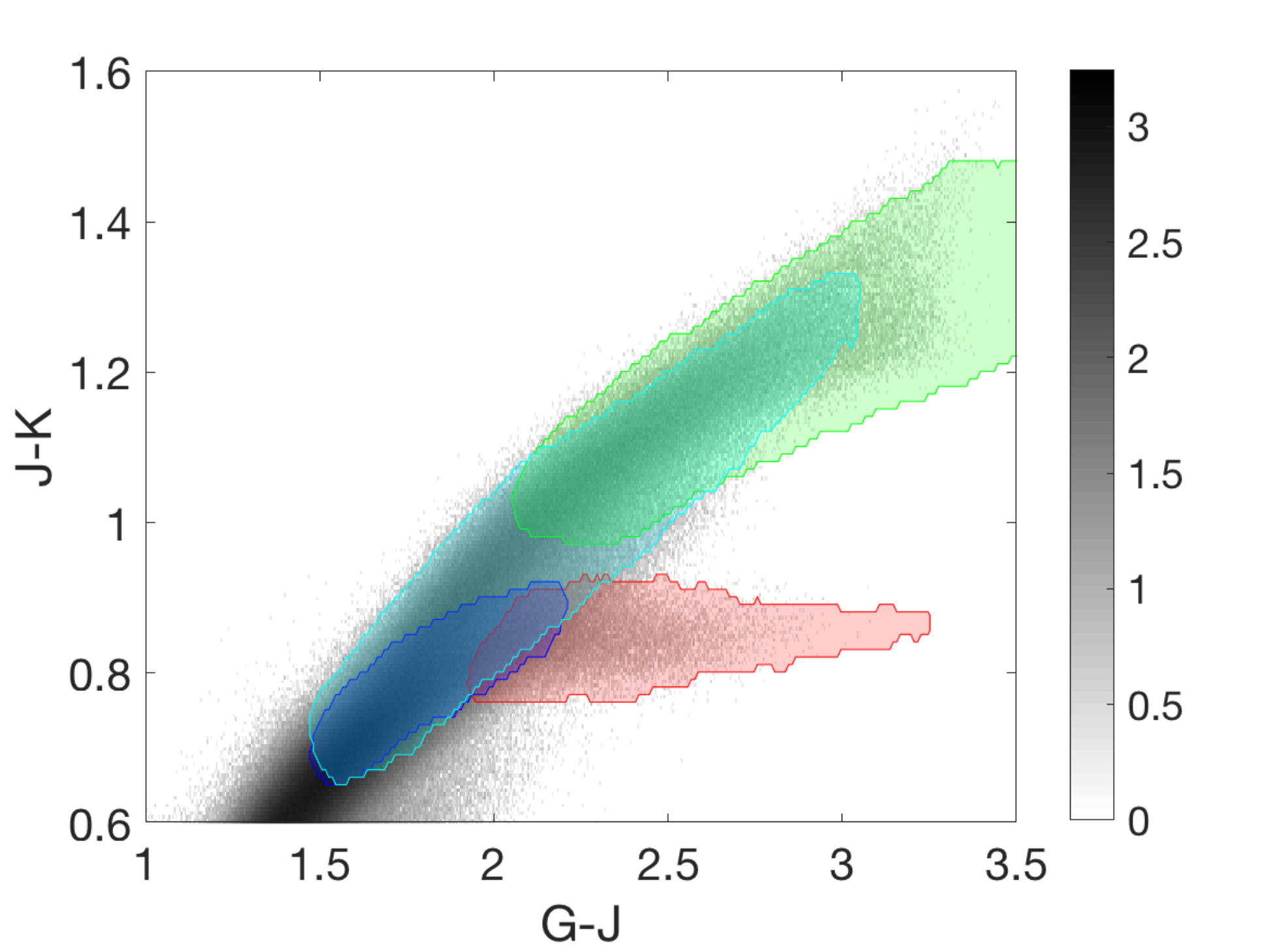}}\\[-1mm]
    \subfloat[]{\label{figColSpaceJKW1W2}\includegraphics[width=0.4\textwidth]{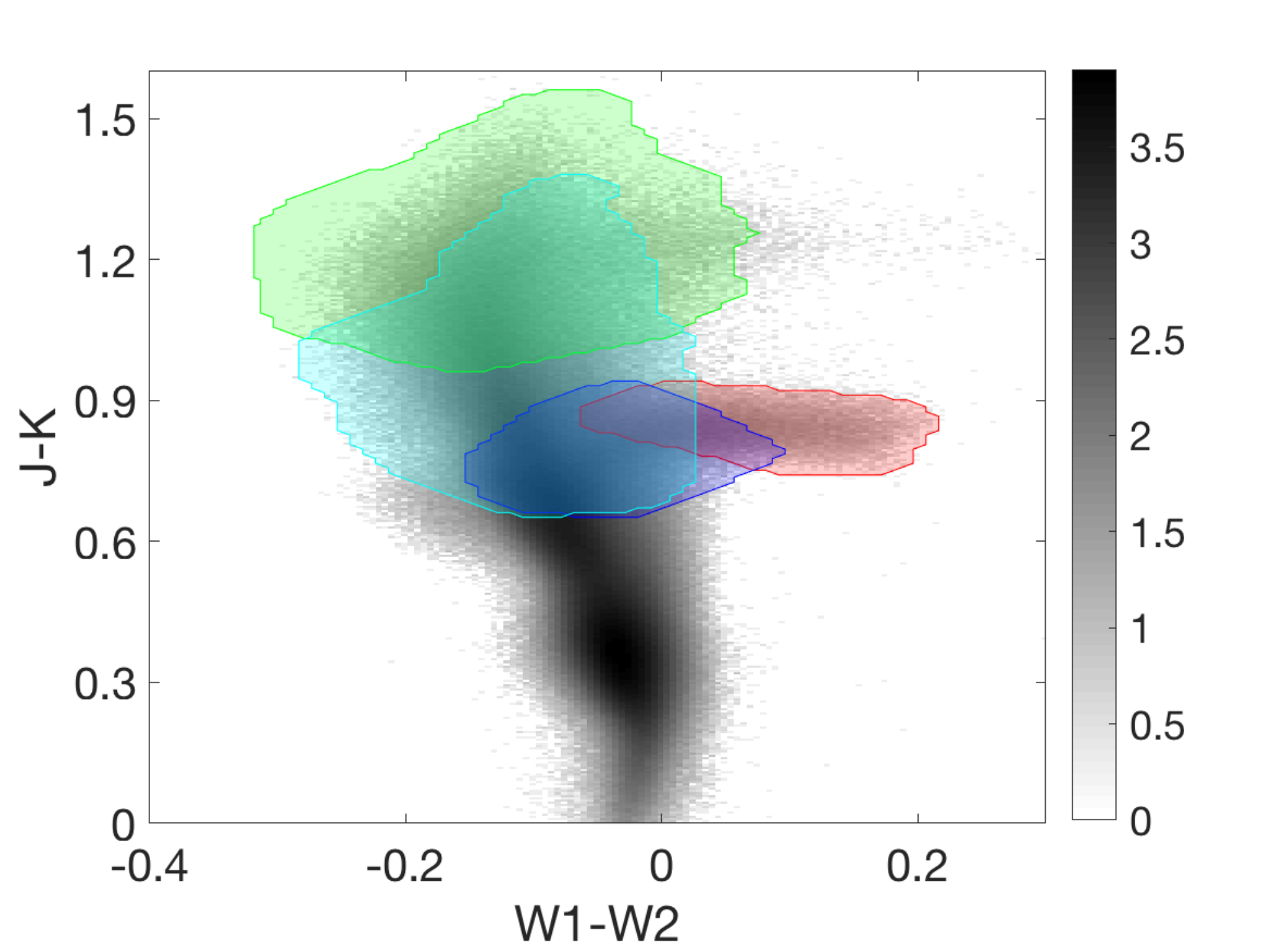}}\\[-1mm]
    \caption{The three colour-colour spaces chosen for this work. Coloured regions highlighting the four classes of stars of importance to this work, as identified by the CCA analysis described in \S\ref{secCCA}. In each case these connected regions contain \textgreater\,98\% of the relevant class in our {\em Galaxia} simulation. Red - M-dwarfs; Blue - K-dwarfs; Cyan - K-giants; Green - M-giants.}
    \label{figCCP}
\end{figure}

{\em G--K/K--W2}: This plane displays the smallest levels of K-dwarf/M-dwarf overlap in all the colour-colour planes  (Figure\,\ref{figColSpaceGKKW2}). In all colour-colour planes, K-dwarfs are the most significant source of contamination in the M-dwarf region, so selecting a colour-colour space that minimises this overlap greatly lowers the overall level of contamination. \\

{\em J--K/G--J}:  This plane has the advantage of using the G-band in the temperature-dependent horizontal axis (Figure\,\ref{figColSpaceJKGJ}) to distribute the M-dwarf branch across a large colour range.\\

{\em J--K/W1--W2}: This has the advantage of using a colour on the horizontal axis for which both bands come from the same survey, telescope and instrument (Figure\,\ref{figColSpaceJKW1W2}), which reduces issues associated with cross-matching, and systematics between surveys with different sensitivities, instrument resolution and data processing.\\

\section{Analysis}
We were guided by the principle of keeping our photometric selection criteria as simple as possible. We first experimented with simple rectangular selection regions in colour-colour space, bounded by the maxima and minima of the CCA regions for M-dwarfs. However, these were quickly found to be unsatisfactory -- there was simply too much contamination at the boundaries between between classes of object, which do not follow either vertical or horizontal lines in these planes.\\

\begin{figure}
	\centering
    \includegraphics[width=0.4\textwidth]{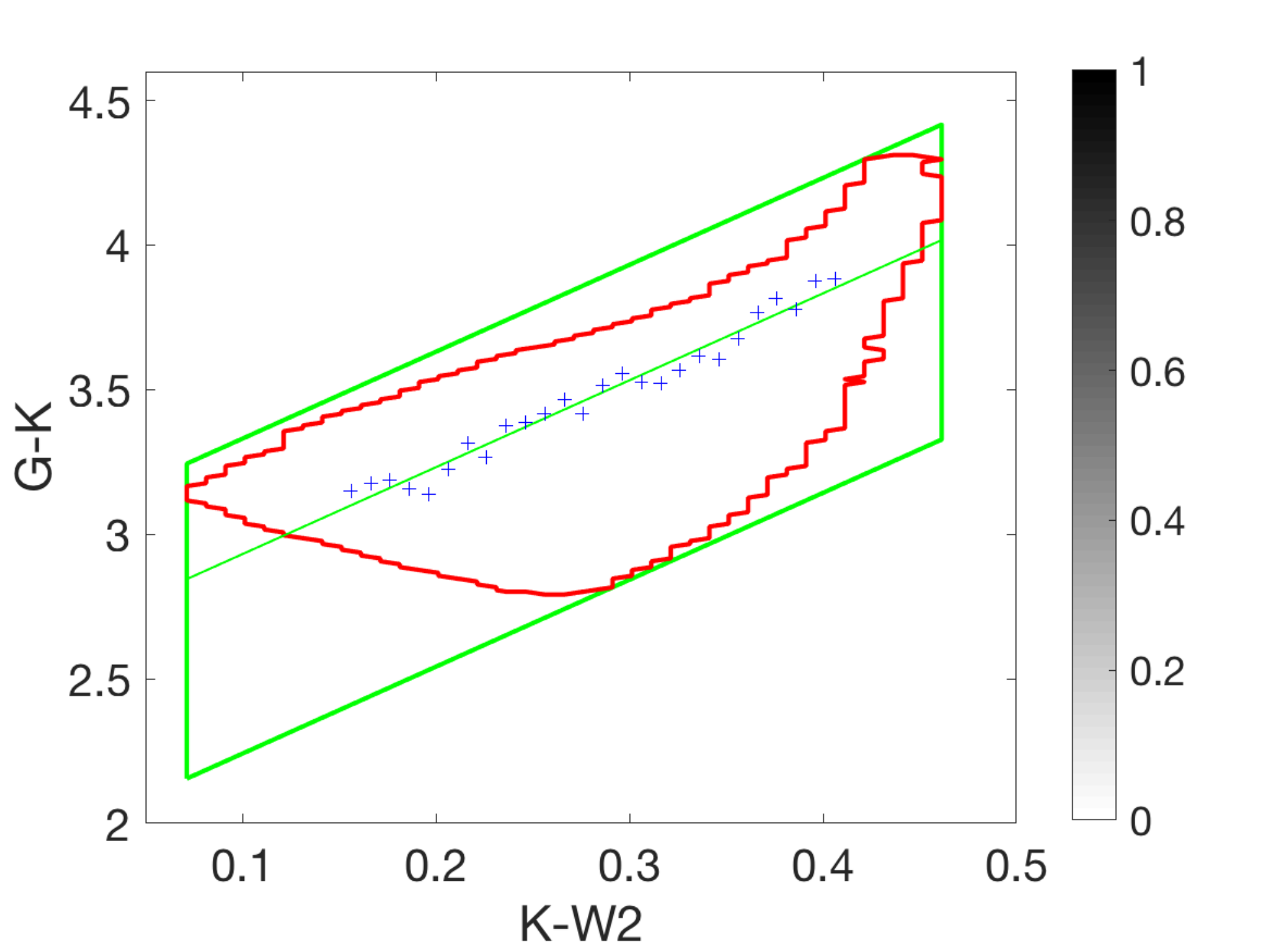}
    \caption{Construction of a parallelogram selection region using the CCA information seen in Figure\,\ref{figCCA}. The red line bounds the primary connected component, the blue crosses represent the G-K values of highest density for the middle 60\% of the K-W2 bins. The thin green line is a linear fit through the blue crosses and the thick green lines represent the parallelogram formed from the intersection of the polynomial with the minimum and maximum K--W2 values from the primary connected component.}
    \label{figPara}
\end{figure}

The next simplest model (i.e. parallelograms) were utilised as the simplest polygon able to represent the M-dwarf region. We binned the data in the M-dwarf CCA region along each horizontal axis and fitted a linear polynomial to the middle 60\% of points in each bin.  Figure\,\ref{figPara} shows an example of this for the data plotted in Figure\,\ref{figCCA}.\\

\begin{figure}
	\centering
    \subfloat[]{\label{figBoxA}\includegraphics[width=0.35\textwidth]{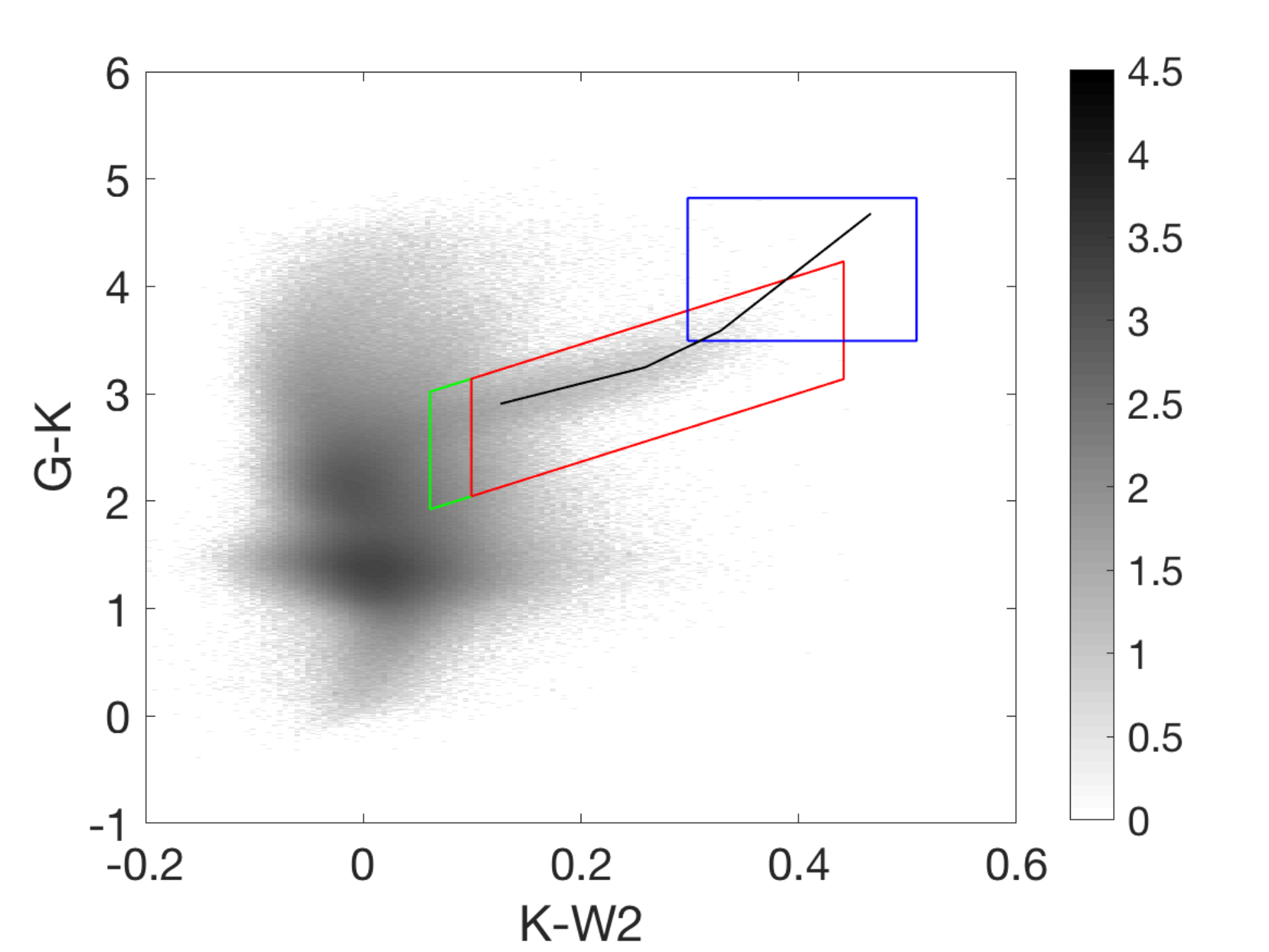}} \\
    \subfloat[]{\label{figBoxB}\includegraphics[width=0.35\textwidth]{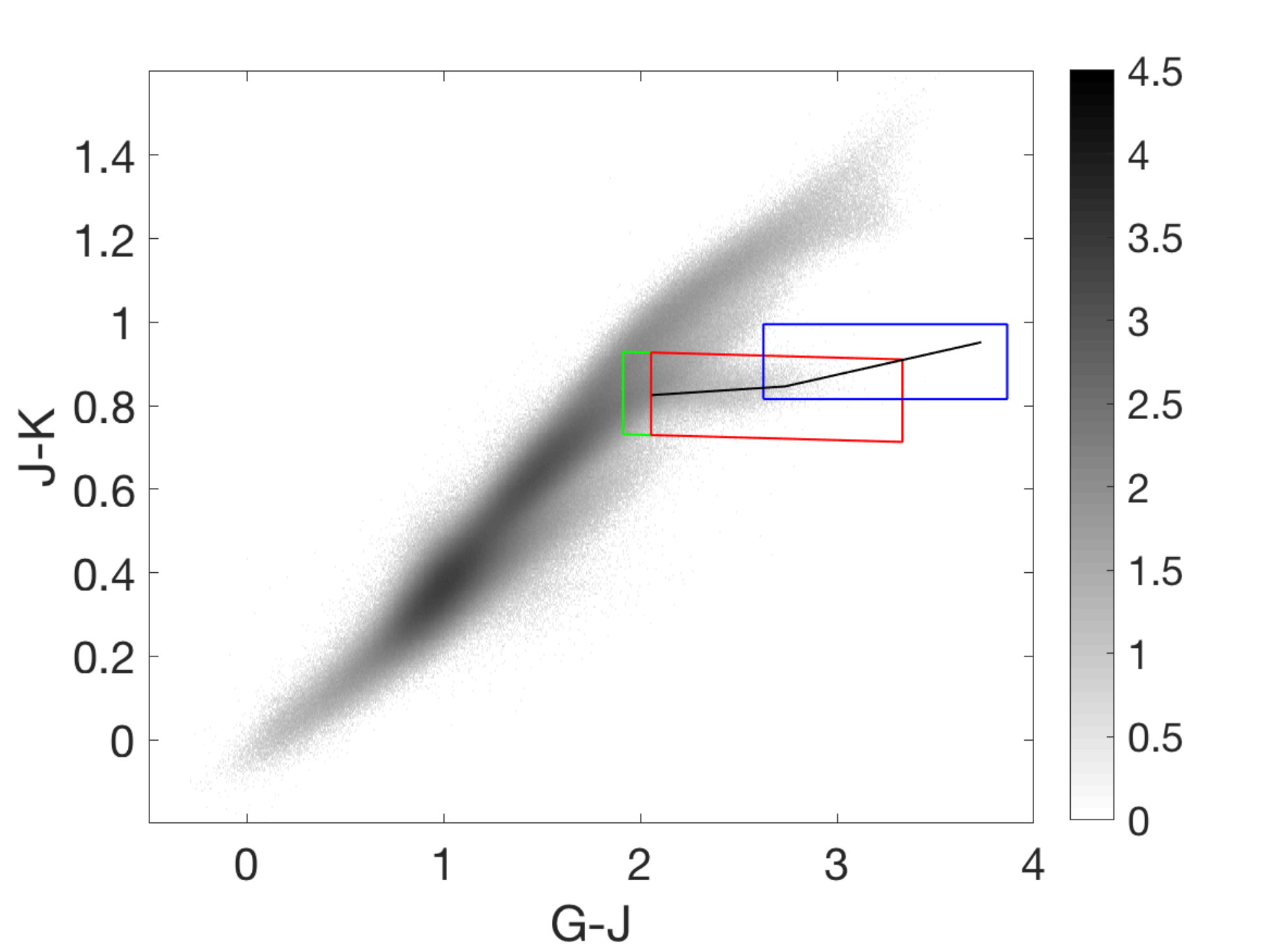}} \\
    \subfloat[]{\label{figBoxC}\includegraphics[width=0.35\textwidth]{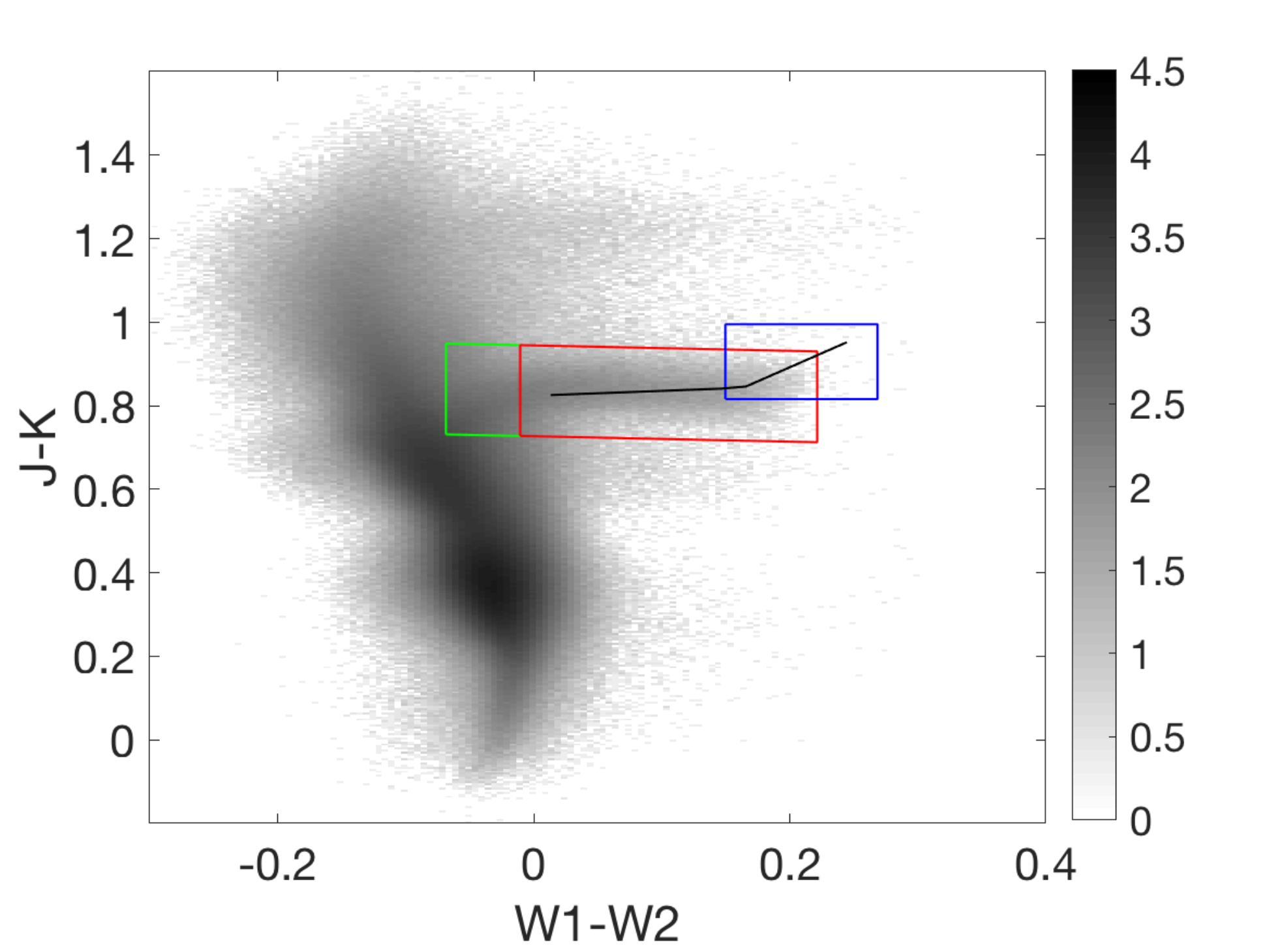}}
    \caption{Colour-colour density plots of {\em Gaia}/WISE/2MASS photometry with the colour regions for the low contamination (in red) and high completeness (in green) criteria for G\,\textless\,14.5. The blue box contains the M6-8 stars that are absent from our low contamination and high completeness colour regions. The black line represents the M-dwarf median colour values from Table\,\ref{TabLookup}.}
	\label{figBox}
\end{figure}

This results in the three groups of ``high completeness'' criteria shown in Equations \ref{eqHC1}-\ref{eqHC3}, which are shown in Figure\,\ref{figBox} overplotted on our observational data as by the combined red and green parallelograms. We then require that an object must satisfy all three of these colour-colour criteria to be considered a candidate M-dwarf.\\

\begin{align}
\hspace{1.5cm}\begin{split}
        K-W2 > 0.061\\
        K-W2 < 0.441\\
        G-K > 3.19(K-W2) + 1.944\\
        G-K < 3.19(K-W2) + 3.019\\
        \label{eqHC1}
    \end{split}
\end{align}
\begin{align}
\hspace{1.5cm}\begin{split}
    	G-J > 1.911\\
    	G-J < 3.331\\
    	J-K > -0.013(G-J) + 0.732\\
    	J-K < -0.013(G-J) + 0.929\\
        \label{eqHC2}
    \end{split}
\end{align}
\begin{align}
\hspace{1.5cm}\begin{split}
		W1-W2 > -0.069\\
    	W1-W2 < 0.221\\
    	J-K > -0.065(W1-W2) + 0.731\\
    	J-K < -0.065(W1-W2) + 0.949\\
        \label{eqHC3}
    \end{split}
\end{align}

While the completeness of this combined set of criteria is very high (99.3\% -- see Discussion), the number of simulated non-M-dwarfs significantly outnumbers the number of simulated M-dwarfs, with most of the non-M-dwarf contamination arising at the transition point from K-dwarfs/giants to M-dwarfs.\\

The conditions that define both sets of criteria consist of four colour limits per colour-colour plane. The most influential is the blue limit of the x-axis as this is where the transition point between K-stars and M-dwarfs occur. This limit will affect both the contamination levels, and what fraction of the total number of M-dwarfs (in particular, M0) are within our selected region. Moving this limit to bluer values will increase the number of M-dwarfs selected, but will add a substantial amount of K-stars. Reducing the size of the selected region by moving this limit to redder values will reduce the K-star contamination, at the expense of the M-dwarf completeness. The other three limits, while important, are not as crucial as the blue x-axis limit. \\

To lower this contamination rate we constructed a second set of ``low contamination'' criteria (Equations \ref{eqLC1}-\ref{eqLC3}), by moving the lower x-axis boundary conditions of each colour region to reduce the number of stars selected. The stars excluded from the colour region will include both M-dwarfs and non-M-dwarfs. The optimal colour region will exclude as many non-M-dwarfs as possible while minimising the amount of M-dwarfs that are lost. We decided to divide the x-axis range of each colour region by twenty, move the lower boundary condition in by that amount and determine the number of M-dwarfs and non-M-dwarfs in the selected region. This was calculated for all three colour spaces and all permeations. The optimal set of colour ranges by maximising the Equation\,\ref{eqFrac}, where $M_{tot}$ is the total number of stars identified in {\em Galaxia} as M-dwarfs, $M_{sel}$ is the number of {\em Galaxia} identified M-dwarfs within our selected colour region and $S_{sel}$ is the total number of stars within our selected colour region.\\

\begin{equation}
\hspace{3cm}f = \frac{M_{sel}}{M_{tot}} \times \frac{M_{sel}}{S_{sel}}
\label{eqFrac}
\end{equation}

The low contamination criterion has non-M-dwarfs comprising $\approx$36\% of the total selected sample, while maintaining M-dwarf completeness at $>$95\%.\\

\begin{align}
\hspace{1.5cm}\begin{split}
        K-W2 > 0.099\\
        K-W2 < 0.441\\
        G-K > 3.19(K-W2) + 2.045\\
        G-K < 3.19(K-W2) + 3.140\\
        \label{eqLC1}
    \end{split}
\end{align}
\begin{align}
\hspace{1.5cm}\begin{split}
    	G-J > 2.053\\
    	G-J < 3.331\\
    	J-K > -0.013(G-J) + 0.730\\
    	J-K < -0.013(G-J) + 0.927\\
        \label{eqLC2}
    \end{split}
\end{align}
\begin{align}
\hspace{1.1cm}\begin{split}
		W1-W2 > -0.011\\
    	W1-W2 < 0.221\\
    	J-K > -0.065(W1-W2) + 0.728\\
    	J-K < -0.065(W1-W2) + 0.945\\
        \label{eqLC3}
    \end{split}
\end{align}

As noted earlier, we expect late M-dwarfs (i.e. later than M5 -- shown in blue in Figure\,\ref{figSubtypes}) to appear off the red end of the {\em Galaxia} M-dwarf branch, because {\em Galaxia} cannot simulate them. This can be seen from Figure\,\ref{figBox} in which our observed M-dwarf photometric sequences are shown as the black curves. To select these late M-dwarfs would necessitate moving the ``vertical'' red cut-off in K--W2, G--J and W1--W2 in Equations\,\ref{eqHC1}-\ref{eqHC3} and \ref{eqLC1}-\ref{eqLC3} to a redder value, as well as also allowing slightly redder colours in G--K and J--K. \\

We expect the number of such very late M-dwarfs to be small. Nonetheless, we developed a supplementary selection criteria to cover the full M-dwarf range, using the median colours and r.m.s. scatter for M6-M8 spectroscopic comparison data (\S\ref{SpecData}). As this area is far from the main sources of contamination (K-stars and giants) using a rectangle to bound this region is adequate, and these late M-dwarf colour regions are indicated by the blue boxes in Figure\,\ref{figBox}.\\

Selecting the full range of M-dwarfs requires the late M-dwarf criterion and either the high completeness or low contamination criterion. As such we define the ``HiComp'' criteria as Equations\,\ref{eqHC1}-\ref{eqHC3} plus \ref{eqLM1}-\ref{eqLM3} and the ``LoCont'' criteria as Equations\,\ref{eqLC1}-\ref{eqLC3} plus \ref{eqLM1}-\ref{eqLM3}.

\begin{align}
\hspace{3cm}\begin{split}
        K-W2 > 0.298\\
        K-W2 < 0.508\\
        G-K > 3.493\\
        G-K < 4.823\\
        \label{eqLM1}
    \end{split}
\end{align}
\begin{align}
\hspace{3.3cm}\begin{split}
    	G-J > 2.625\\
    	G-J < 3.864\\
    	J-K > 0.816\\
    	J-K < 0.995\\
        \label{eqLM2}
    \end{split}
\end{align}
\begin{align}
\hspace{2.9cm}\begin{split}
		W1-W2 > 0.150\\
    	W1-W2 < 0.268\\
        \label{eqLM3}
    \end{split}
\end{align}

\section{Discussion}
\subsection{Colour Selection pre-{\em Gaia} DR2}

\subsubsection{Completeness and Contamination in Simulations}

Our simulated data allow us to quantify both the completeness and  contamination (i.e. the false-positive rate) for M-dwarf candidate selection using these criteria. We define simulated completeness as the number of M-dwarfs selected, divided by the total number of M-dwarfs present in the simulation, and simulated contamination as the number of non-M-dwarfs identified by each set of criteria, divided by the total number of object selected by each set of criteria. These results are reported in Table \ref{tabCrit}.\\

The HiComp criterion (Eqs \ref{eqHC1}-\ref{eqHC3} and \ref{eqLM1}-\ref{eqLM3}) is designed to maximise the selection of M-dwarfs, at the cost of higher non-M-dwarf contamination. We predict that this criterion would select 99.3\% of all M-dwarfs, at the cost of 55.6\% of the objects selected not actually being M-dwarfs. Whether this level of contamination is acceptable for M-dwarf selection will depend greatly on the total number of candidates identified and the resources available for follow-up of those candidates. If the total number identified is $\sim$50,000 objects and they can be observed spectroscopy as part of a survey targeting 1,000,000 stars a year (as is the goal of the {\em FunnelWeb} survey), then the cost of a contamination rate of 55.6\% is perfectly acceptable. If they are being followed-up by observation one-at-a-time, then that cost would be prohibitive.\\

The LoCont criterion (Eqs \ref{eqLC1}-\ref{eqLC3} and \ref{eqLM1}-\ref{eqLM3}) reduces the number of M-dwarf candidates by about a third down to 37,079 giving a contamination rate of just 35.9\%, at the cost of lowering M-dwarf completeness to 98\% (i.e. missing ~300 simulated M-dwarfs, almost all of which will be early M0 types).

\subsubsection{Contamination in Observations}

Applying the same selection criteria to our observational sample selects the number of objects that ``Meet Criteria'' listed in Table \ref{tabCrit}. The first point that is apparent from these numbers is that the number of observed objects that meet the HiComp criterion is around 44\% higher than the number predicted by the {\em Galaxia} simulation -- i.e. 78,340 vs 54,255. The corresponding difference for the LoCont sample is much smaller at 9.2\% (40,505 vs 37,079).\\

As noted earlier (\S\ref{SecNorm}), it was found to be problematic to obtain a single normalisation between our {\em Galaxia} and observational data sets, with different regions of the simulated colour-colour plane requiring different normalisations. Our chosen overall normalisation was one based on the population of main sequence stars with J--W1\,\textless\,0.67, where {\em Galaxia} has been most tuned to accurately match the observed Galaxy. \\

There are two ways to look at the difference between the total number of objects selected by our criteria in our simulated data and in observed data. At one extreme, the number of simulated M-dwarfs could be correct and the simulations fail to reproduce the way the real world scatters non-M-dwarfs into the colour selection criteria regions. At the other extreme the simulations could be correctly predicting the total number of objects in the colour selection regions, and under-predicting the total M-dwarf stellar population -- in this situation, our best estimate of the number of M-dwarfs to be identified will be the number of observed objects that meet our criteria, times one-minus-the-contamination-rate derived from our simulations. The fact that the difference between the predicted and observed number counts is significantly smaller for the LoCont sample (which digs into the K-dwarf, K-giant and M-giant regime much less than the HiComp sample) suggests the former is more likely. However we cannot rule out the latter.\\

We therefore provide in Table\,\ref{tabCrit} two estimates for the total number of M-dwarfs predicted to be found in the observed sample. They are the extreme bounds that arise from these two assumptions. For the HiComp criterion the contamination rate range is 55.6-69.3\%, while for the LoCont criterion it is 35.9-41.3\%.\\

The key point here is that the total numbers of objects predicted ($\approx$78k and $\approx$40k for the two sets of criteria respectively) are {\em both} tractable numbers for a massively multiplexed, all-sky spectroscopic survey like {\em FunnelWeb}, where the cost to observe $\sim$100,000 objects a year is low, as the time scale is relatively short. In this situation, obtaining a highly complete spectroscopic sample becomes a clear priority.\\

\subsubsection{Impact of the latest M-dwarfs}

{\em Galaxia} (as noted earlier) cannot predict the number of late M-dwarfs we will find (or miss) for these photometric selection criteria. However, we can look at our observational data to at least estimate the scale of the problem. \\

If we look at the number of objects selected by Equations\,\ref{eqHC1}\,-\,\ref{eqHC3} plus \ref{eqLM1}\,-\,\ref{eqLM3} vs \ref{eqHC1}\,-\,\ref{eqHC3} alone we find 249 extra objects in our observational data that are potential late-M-dwarfs not captured by our early-M-dwarf criterion. (This number is the same for both the HiComp and LoCont criteria). As such, the inclusion of late M-dwarf candidates comes at almost no cost to the {\em FunnelWeb} survey, making their inclusion eminently sensible.\\

\begin{table*}
	\begin{tabular}{ | c | c c c c | c c c | } 
		\hline
		& \multicolumn{4}{c|}{Simulations} & \multicolumn{3}{c|}{Observations}\\
		\cline{2-8}
		Criteria & Meet & Total & Completeness & Contamination & Meet     & \multicolumn{2}{c|}{Predicted M-dwarfs}\\
		         & Criteria & M-dwarfs & (\%) & (\%)           & Criteria & (Number)  & (\%)\\
		\hline
		HiComp & 54255 & 24078 & 99.3 & 55.6 & 78340 & 24078\,-\,34767 & 30.7\,-\,44.4\\
		LoCont & 37079 & 23764 & 98 & 35.9 & 40505 & 23764\,-\,25960 & 58.7\,-\,64.1\\
		\hline
	\end{tabular}
    \caption{Simulation and Observation predictions for M-dwarf selection using High Completeness (HiComp) and Low Contamination (LoCont) criteria. ``Meet Criteria'' numbers are all objects in either samples that meet the HC and LC criteria. The simulated ``Total M-dwarf'' numbers are simulated objects identified as M-dwarfs that are selected by each set of criteria, and ``Completeness'' is the percentage of that number of the total number of simulated M-dwarfs. ``Contamination'' is the percentage of simulated objects that ``Meet Criteria'' but are not simulated M-dwarfs. ``Predicted M-dwarfs'' are the extreme ranges for numbers of objects in the Observational sample that ``Meet Criteria'' and could be real M-dwarfs (see text). }
    \label{tabCrit}
\end{table*}

\subsection{Absolute Magnitude Selection post-{\em Gaia} DR2}

In an ideal world, M-dwarf selection would be made on the basis of the one parameter that clearly distinguishes them from all other classes of stars -- luminosity (with perhaps a simple colour selection required to distinguish degenerate white dwarfs from main sequence red dwarfs). Our exploration of M-dwarf selection using photometric colour criteria was primarily motivated by the desire to perform a kinematically unbiased selection for M-dwarfs in the absence of the distances required to make a purely absolute magnitude-limited selection. \\

However, with the release of {\em Gaia} DR2, international astronomy will enter into that ``ideal'' world. We have therefore used our simulated data to examine the impact of such a revolutionary set of complete, all-sky distance measurements.\\

\begin{figure}
	\centering
    \subfloat[]{\label{figColMagA}\includegraphics[width=0.4\textwidth]{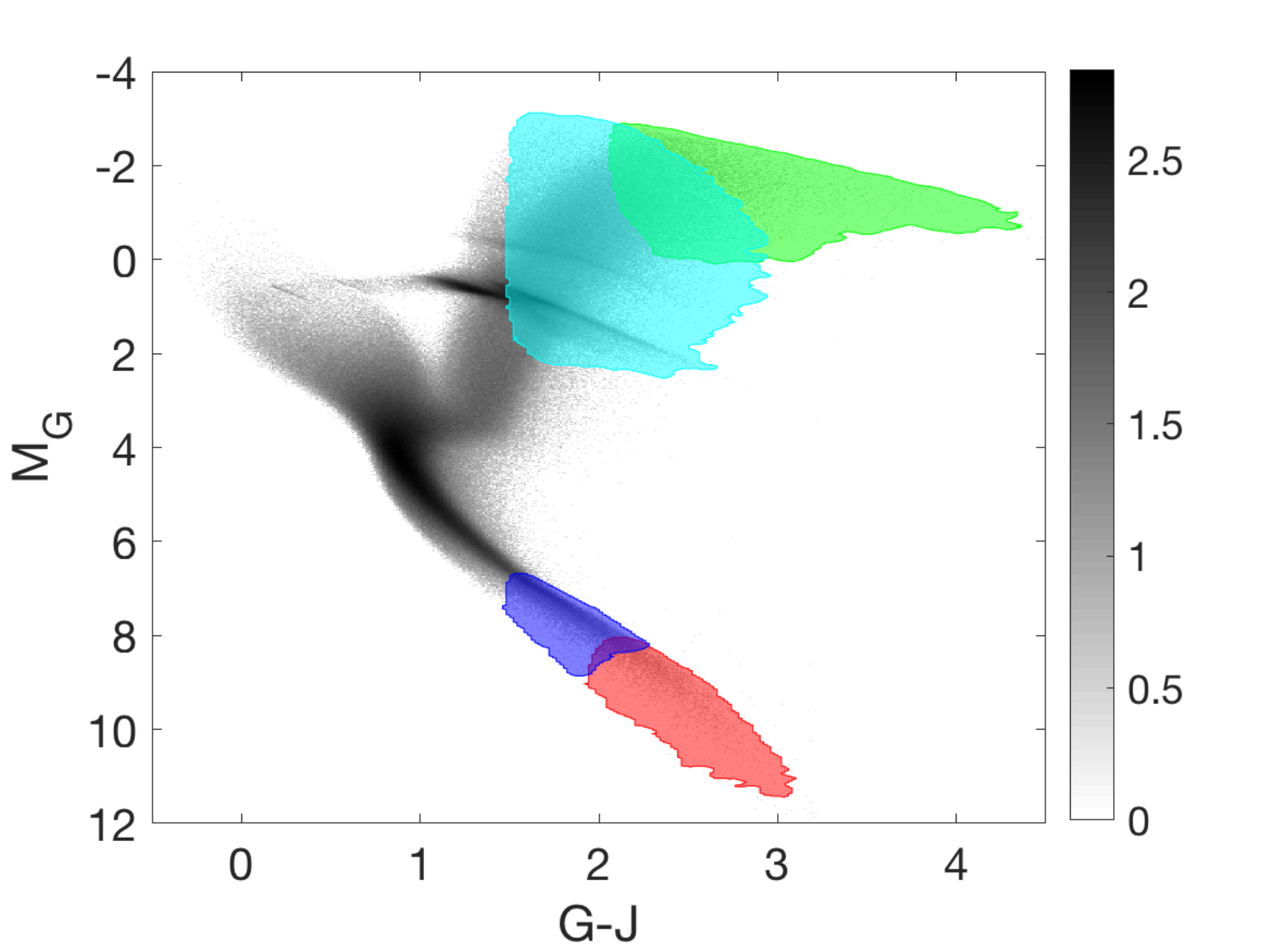}} \\
    \subfloat[]{\label{figColMagB}\includegraphics[width=0.4\textwidth]{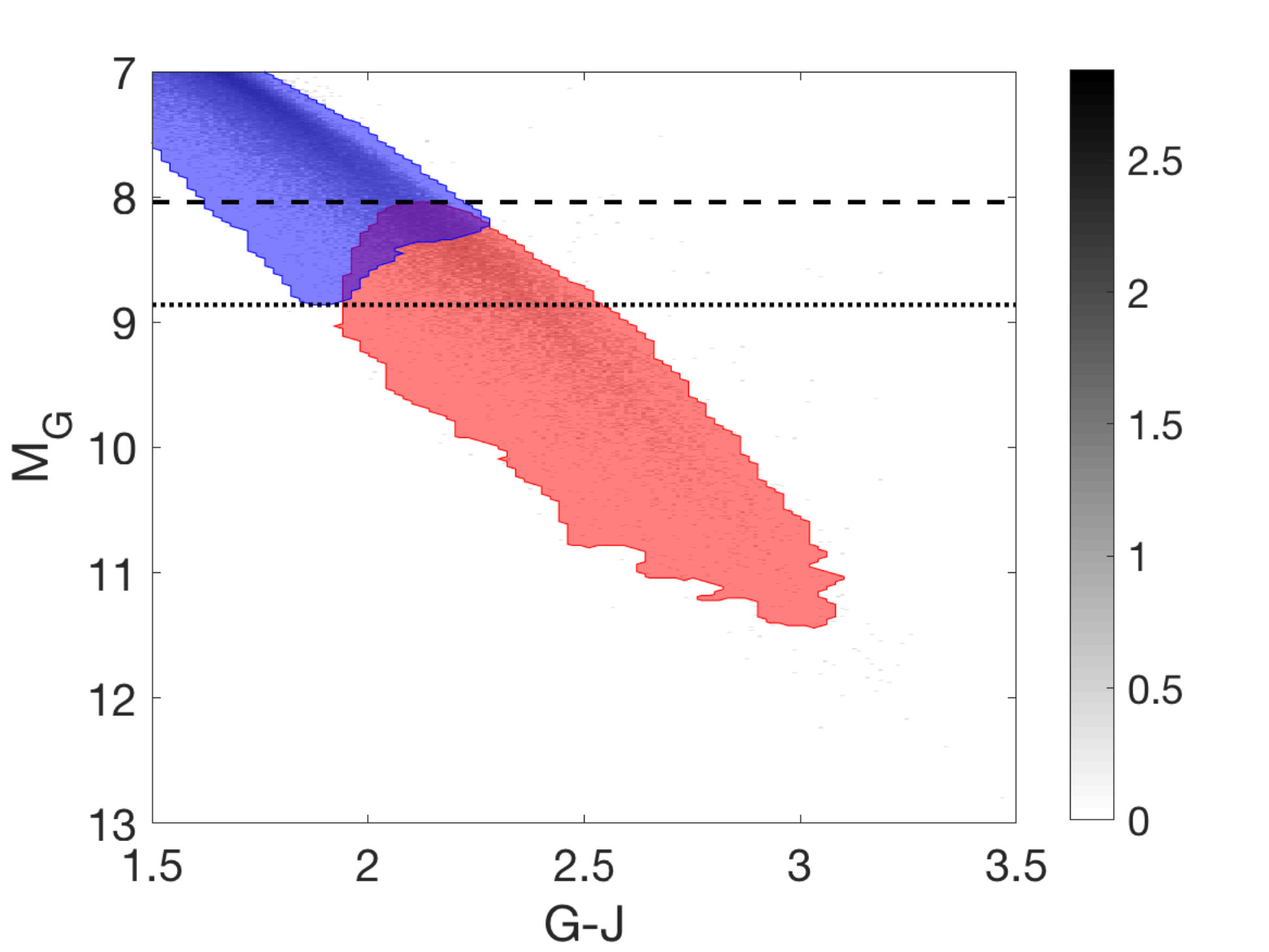}} \\
    \caption{Simulated M$_G$:G--J colour magnitude diagram for G\,\textless\,14.5. The K and M, dwarf and giant regions are identified using the criteria of Table\,\ref{TabMK}, and the overplotted CCA regions (complete to \textgreater\,97\%) use the same colour scheme as in Figure\,\ref{figCCP}. \ref{figColMagB} is a zoomed in image of the K to M dwarf boundary seen in Figure\,\ref{figColMagA} with black horizontal lines representing the absolute magnitude limits of HiCompAbs and LoContAbs.}
    \label{figAbs}
\end{figure}

We calculated the absolute magnitude M$_G$ for all our simulated targets, and used them to construct a representative M$_G$:G--J colour-magnitude diagram (Figure\,\ref{figColMagA}), and then used the characteristics from Table\,\ref{TabMK} and CCA techniques to identify and select the M and K dwarf, and M and K giant populations in this plane. It is clear that the issue of contamination of an M-dwarf sample by M and K giants is completely removed, and the only issue in such a plane is the degree of K dwarf contamination as a function of an adopted high luminosity limit in M$_G$ for M-dwarfs.\\

Informed by these CCA regions, we have chosen two M$_G$ cut-offs for M dwarf selection. A ``high completeness'' absolute magnitude (HiCompAbs) limit, and a ``low contamination'' absolute magnitude (LoContAbs) limit, as indicated in Equations\,\ref{eqHiCompAbs} and\ref{eqLoContAbs} and seen as the dashed and dotted black lines in Figure\,\ref{figColMagB}.
\begin{align}
\hspace{3cm}\begin{split}
        M_G > 8.04\\
        \label{eqHiCompAbs}
    \end{split}
\end{align}
\begin{align}
\hspace{3cm}\begin{split}
    	M_G > 8.86\\
        \label{eqLoContAbs}
    \end{split}
\end{align}

These absolute magnitude limits were deliberately chosen to be analogous to our HiComp and LoCont colour selection criteria. Table\,\ref{tabAbs} shows the same simulated completeness and contamination rates as shown for our colour selection criteria in Table \ref{tabCrit}. Using HiCompAbs M$_G$ selection in isolation, significantly reduces the contamination rate of  non-M-dwarfs (compared to the colour HiComp selection) by a factor of about 2. Moreover it does so while maintain M-dwarf completeness at 99.96\%, a level higher than the HiComp criterion. Perhaps most critically, the HiCompAbs criterion selects more candidate M-dwarfs than the HiComp and LoCont colour selection criteria. (The LoContAbs selection would achieve an unacceptably poor level of completeness, so we do not consider it further).

Because the fraction of observed stars with pre-existing {\em Gaia} DR1 astrometry is so small ($\sim$12.5\% of the stars in our {\em Gaia}/WISE/2MASS dataset have parallaxes from the Tycho-Gaia Astrometric Solution; \citealt{2015Michalik}), we are unable to compare these predictions with the real world in any meaningful fashion, until the release of {\em Gaia} DR2. Nonetheless it is both clear (and unsurprising) that precision distances will make a significant change in our ability to select M-dwarfs across the whole sky. \\

\begin{table*}
	\begin{tabular}{ | c | c c c c | } 
		\hline
		& \multicolumn{4}{c|}{Simulations}\\
		\cline{2-5}
		Criteria & Meet & Total & Completeness & Contamination\\
		         & Criteria & M-dwarfs & (\%) & (\%)\\
		\hline
		HiCompAbs & 34324 & 24247 & 99.96 & 29.4\\
		LoContAbs & 15963 & 13277 & 54.7 & 16.8\\
		\hline
	\end{tabular}
    \caption{Simulated predictions for M-dwarf selection using an absolute magnitude cut of M$_G$\,=\,8.04 for HiCompAbs and M$_G$\,=\,8.86 for LoContAbs.}
    \label{tabAbs}
\end{table*}

\subsection{Absolute Magnitude plus Colour Selection post-{\em Gaia} DR2}

We now consider the impact of requiring both the colour and absolute magnitude criteria to jointly apply - i.e. the HiComp plus HiCompAbs criteria (hereafter, HiComp+), and the LoCont plus LoContAbs criteria (LoCont+). The results of such a selection are shown in Table \ref{tabCritABs} in the same format as Table \ref{tabAbs}.\\

\begin{table*}
\begin{tabular}{ | c | c c c c | } 
		\hline
		& \multicolumn{4}{c|}{Simulations}\\
		\cline{2-5}
		Criteria & Meet & Total & Completeness & Contamination\\
		         & Criteria & M-dwarfs & (\%) & (\%)\\
		\hline
		HiComp+ & 32351 & 24068 & 99.2 & 25.6\\
		LoCont+ & 15331 & 13083 & 53.9 & 14.7\\
		\hline
	\end{tabular}
    \caption{Simulated predictions for M-dwarf selection using the HiComp and LoCont colour criteria and an absolute magnitude cut of M$_G$\,=\,8.04 for HiComp and M$_G$\,=\,8.86 for LoCont.}
    \label{tabCritABs}
\end{table*}

\begin{figure}
	\centering
    \subfloat[HiComp v.s. HiComp+]{\label{figMag1}\includegraphics[width=0.4\textwidth]{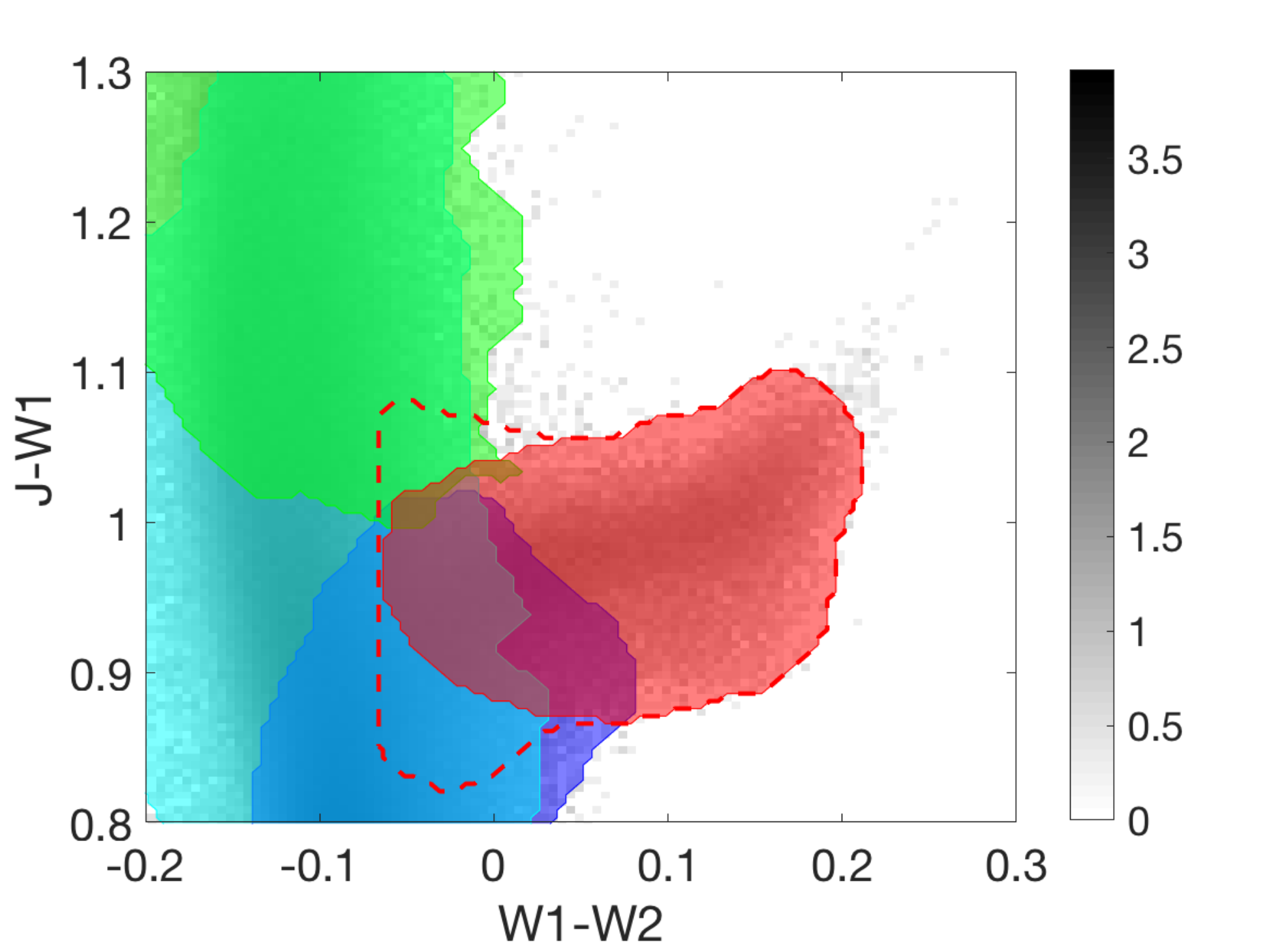}} \\
    \subfloat[LoCont v.s. LoCont+]{\label{figMag2}\includegraphics[width=0.4\textwidth]{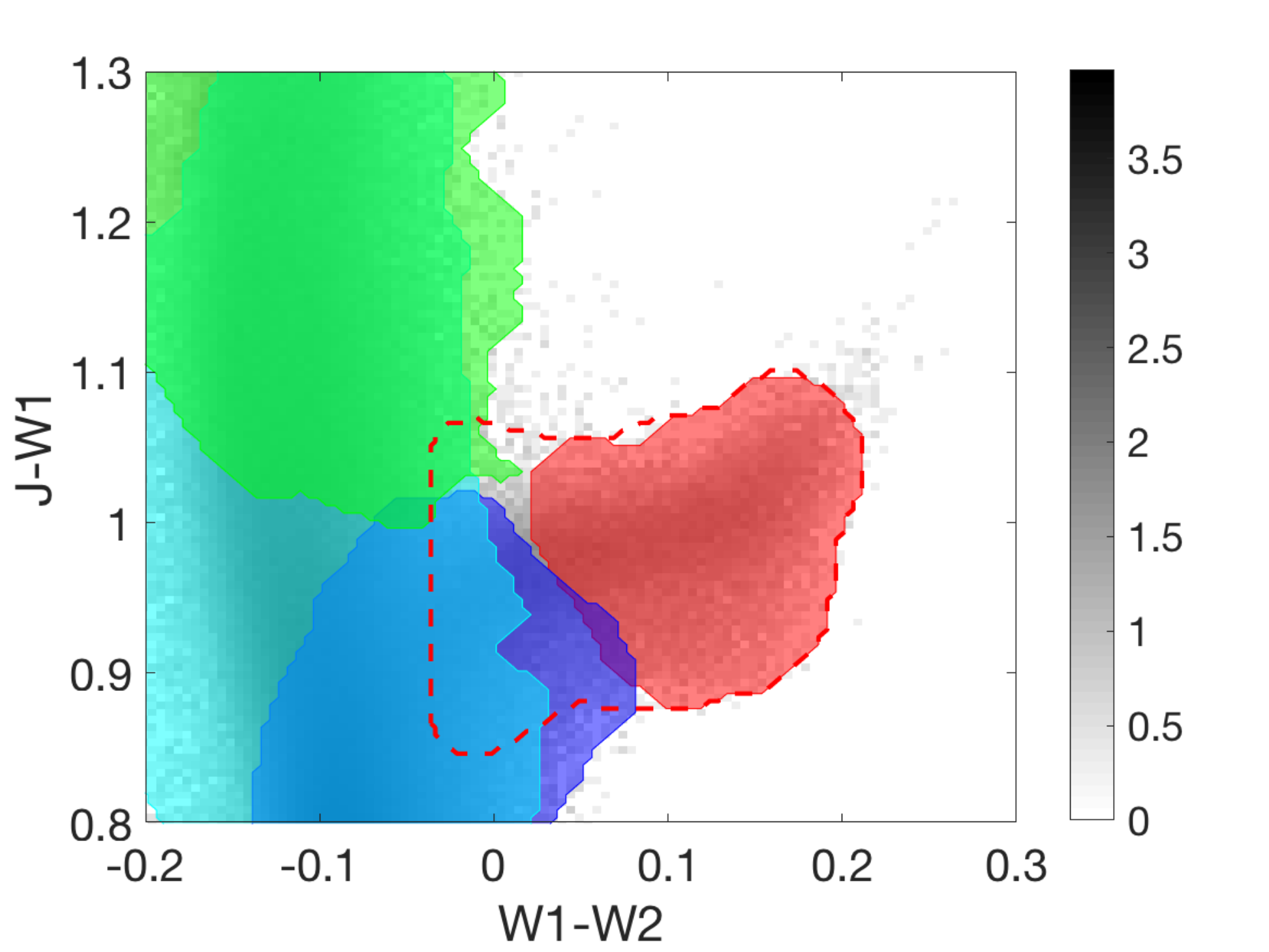}} \\
    \caption{The K-stars and M-giants plus the stars selected by our colour criteria and absolute magnitude cut. The dashed red contour represents the colour region selected by our colour criteria alone (i.e. HiComp and LoCont), while the red shaded area indicates the stars that are selected through colour and absolute magnitude (i.e. HiComp+ and LoCont+). The K-dwarf, K-giant and M-giant regions represent \textgreater\,97\% of their total respective populations and use the same colour scheme as Figure\,\ref{figCCP}.}
    \label{figMagCut}
\end{figure}

The combination of all these criteria both completely removes contaminating giants from an M-dwarf candidate sample, and removes a substantial number of contaminating K-dwarfs. This can be seen more clearly in Figure\,\ref{figMagCut}, which compares the two sets of selections and how far they reach into the various contaminate star regions. The stars included in the HiComp selection but excluded from the HiComp+ selection (Figure\,\ref{figMag2}) are predominantly K-stars (cyan and blue), with a smaller amount of M-giants (green). The numbers of objects selected in Table\,\ref{tabCritABs} reinforce the idea that, once absolute magnitudes are available, the selection of a highly complete (i.e. $>$\,99\%) M-dwarf sample becomes doable, with the requirement to follow-up a small number ($\approx$33k) of targets.\\

The LoCont+ (Figure\,\ref{figMag1}) selection excludes essentially all giants, with the contamination expected to consist of a small amount of K-dwarfs. However, this requires a selection limit so far into the M-dwarf branch to avoid the giant stars and this excludes a high amount of early M-dwarfs, dropping the completeness to just over 50\%! Once again the low contamination option still has sufficient contamination and low enough completeness, as to not be worth utilising.\\

\subsection{Comparison with the TESS Input Catalogue}
\label{secTIC}
\begin{figure}
	\centering
    \includegraphics[width=0.4\textwidth]{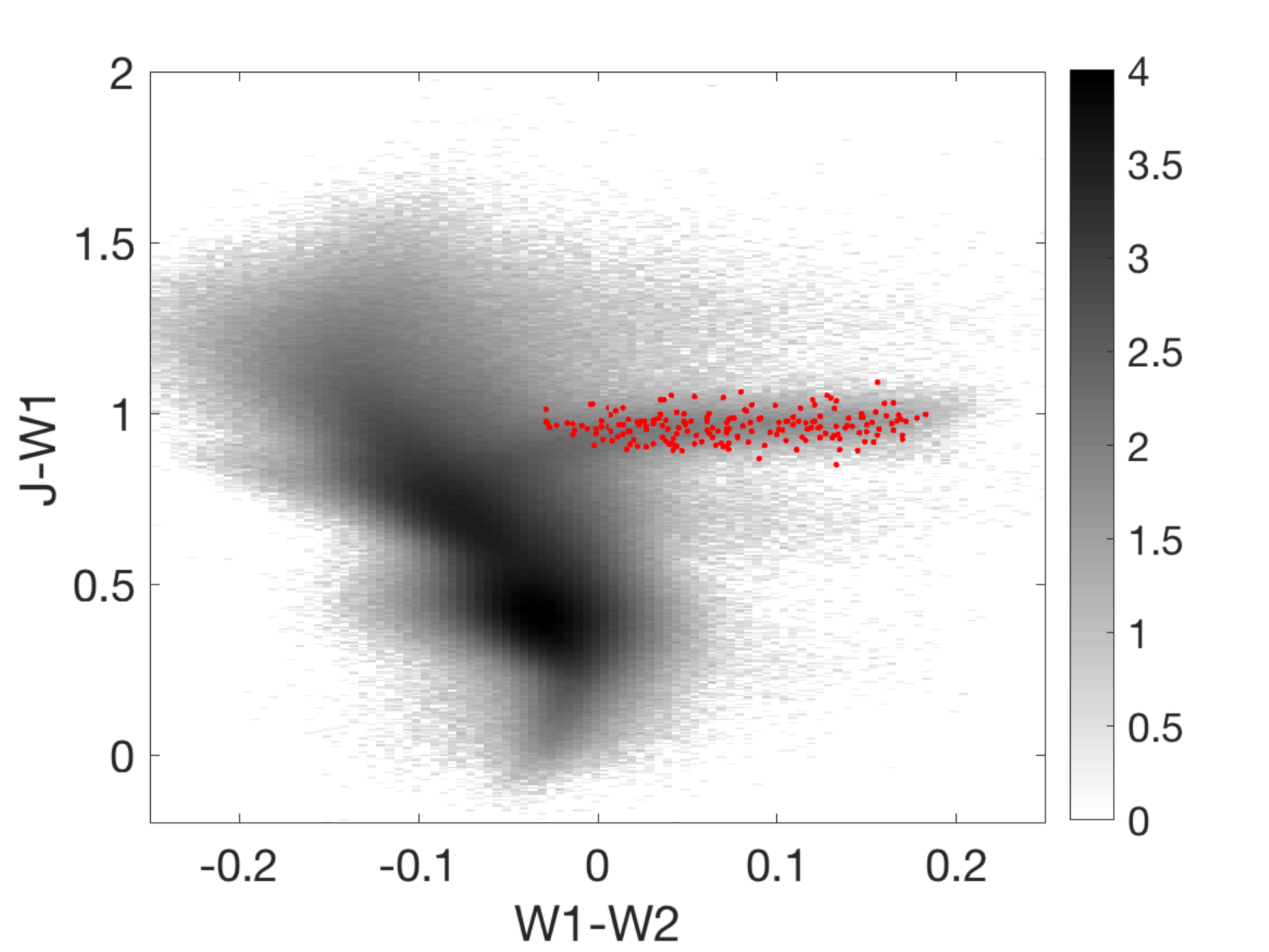}
    \caption{Colour-colour density plot of the observational {\em Gaia}/WISE/2MASS data and the stars that this work has identified as suitable candidates for the TIC, in red.}
    \label{figTIC}
\end{figure}

We have cross-matched our identified HC candidates against version 5 of the TESS Input Catalogue (TIC; \citealt{2017Stassun}). The vast majority of our southern candidates are present in the TIC -- we find only 234 HC sample stars that are not in the TIC. Details for these 234 objects are tabulated in Table\,\ref{TabTIC}, and overplotted in  Figure\,\ref{figTIC}. At the time of writing, the TIC does not contain any classification other than whether the object is a star, galaxy or is unknown. Additionally these 234 stars have no spectroscopy (and so no known spectral class) so we have used the empirical relationships between spectral type and photometric colours in Table\,\ref{TabLookup} to estimate crude spectral types from each of their G--J, G--K, K--W2 and W1--W2 colours. (J--K was not used as the relationship is essentially flat for K7-M4 dwarfs, making it a very poor estimator for early M-dwarfs). As we saw in Figure\,\ref{FigRelationship}, the accuracy with which we can estimate a star's subtype is strongly dependent on the colour and subtype. W1--W2 and K--W2 type estimates are accurate to $\pm$2 subtypes, while G--J and G--K have similar accuracies for stars M5 and later, but are much more imprecise for earlier stars.\\

The mean of the four types indicated by these four colours is shown in Table\,\ref{TabTIC}, and is estimated to be good to $\pm$1.5 sub-types. All of these 234 candidates fall within the colour ranges expected for late K-dwarfs and early M-dwarfs, with approximately 70\% of them expected to be early M-dwarfs (M0-M2 -- see Table\,\ref{TabStats}). The majority of these stars therefore fall within TESS' target range for M-dwarf candidates of K5-M5.\\

\begin{table}
	\centering
	\begin{tabular}{|c|c|c|}
    \hline
    Subtype & \# & \% \\
    \hline
    K7 & 23 & 9.83\\
    M0 & 73 & 31.20\\
    M1 & 55 & 23.50\\
    M2 & 36 & 15.38\\
    M3 & 34 & 14.53\\
    M4 & 13 & 5.56\\
    \hline
    \end{tabular}
	\caption{Distribution of approximate classifications for the 234 HiComp M-dwarf candidates not currently in the TESS Input Catalogue (version 5).}
    \label{TabStats}
\end{table}
\section{Conclusions}
We have developed a set of criteria based on the comparison of synthetic and observational photometry. One criterion (HiComp) focusses on selecting as many M-dwarfs as possible, regardless of the levels of non-M-dwarfs. While the contamination levels in the HiComp criterion are not trivial, multi-star spectroscopic surveys, such as {\em FunnelWeb} will be able to observe all candidate stars in a relatively short period of time and spectroscopically confirm the M-dwarfs.\\

The other criterion (LoCont) is a subset of the first, that reduces the number of non-M-dwarfs, but in doing so, excludes significant levels of early M-dwarfs and as such, is most suitable as a list of high priority candidates on ``one target at a time'' spectroscopic surveys. \\

Once the Gaia DR2 is made publicly available, the best method for the identification of determine M-dwarfs will be to use the calculated distances and fluxes to determine luminosities. Our work shows that the rates of M-dwarf completeness and non-M-dwarf contamination are tractable with colour selection alone, but that absolute magnitude selection is even better. In the post-Gaia DR2 era, the combination of absolute magnitude and colour selection will maximise M-dwarf completeness and minimise contamination.\\
\section*{Acknowledgements}
This work has made use of data from the European Space Agency (ESA) mission \textit{{\em Gaia}} (https://www.cosmos.esa.int/gaia), processed by the \textit{{\em Gaia}} Data Processing and Analysis Consortium (DPAC, https://www.cosmos.esa.int/web/gaia/dpac/consortium). Funding for the DPAC has been provided by national institutions, in particular the institutions participating in the \textit{{\em Gaia}} Multilateral Agreement.\\

\bibliographystyle{mnras}
\bibliography{mnras} 

\appendix
\section{Candidate M-dwarfs for the TESS Input Catalogue}
\label{secAppend}
\onecolumn
\begin{longtable}{| c | c | c | c | c |}
\hline
$\alpha$ (deg) & $\delta$ (deg) & 2MASS & Gaia & Sp. Type \\
IRCS 2000 & IRCS 2000 & Identifier & Identifier & \\
\hline
\endhead

\hline
\caption{Southern hemisphere M-dwarf candidates not currently in the TESS Input Catalogue. The spectral type was estimated using photometry from WISE and a set of classified stars from \citet{2011West} and \citet{2015Zhong}.}
\endfoot

\hline
\caption{Southern hemisphere M-dwarf candidates not currently in the TESS Input Catalogue. The spectral type was estimated using photometry from WISE and a set of classified stars from \citet{2011West} and \citet{2015Zhong}.}
\endlastfoot
1.4272 & -64.1394 & 00054261-6408215 & 4901148088120955264 & M1 \\
1.4302 & -61.1742 & 00054318-6110264 & 4905603858992341888 & M4 \\
1.4399 & -60.988 & 00054565-6059161 & 4905658044300034560 & M2 \\
1.5429 & -60.141 & 00061044-6008267 & 4905941168543855872 & M0 \\
1.5819 & -65.8428 & 00061920-6550262 & 4899957901143352576 & M0 \\
1.6412 & -62.8692 & 00063384-6252087 & 4904458889430313088 & M3 \\
1.6431 & -48.7531 & 00063428-4845109 & 4977749791917230848 & M2 \\
1.7037 & -52.9209 & 00064882-5255157 & 4972113145557887360 & M1 \\
1.7809 & -48.9119 & 00070740-4854431 & 4976993087399245568 & K7 \\
1.8475 & -48.1048 & 00072340-4806162 & 4977838886719223424 & M1 \\
1.8854 & -50.7964 & 00073253-5047468 & 4973546221525747328 & M3 \\
1.9568 & -48.1025 & 00074958-4806089 & 4977836137940154624 & M0 \\
3.1762 & -61.0898 & 00124220-6105215 & 4904941162718314368 & M3 \\
3.2336 & -60.2659 & 00125618-6015561 & 4905786206123597696 & M1 \\
3.2353 & -60.9572 & 00125649-6057259 & 4904945766923326336 & K7 \\
3.2723 & -65.0349 & 00130528-6502056 & 4900134853796249216 & M3 \\
4.3631 & -48.5491 & 00172720-4832568 & 4977166707158065152 & M0 \\
4.3879 & -50.5431 & 00173322-5032347 & 4973717367382150912 & M4 \\
4.3964 & -48.976 & 00173503-4858329 & 4977083934547993088 & M0 \\
4.408 & -50.8909 & 00173783-5053267 & 4972946575371823104 & M0 \\
4.4405 & -52.3699 & 00174586-5222113 & 4972549754753146624 & M1 \\
4.4488 & -51.325 & 00174762-5119305 & 4972873114251308544 & M3 \\
4.7228 & -48.8425 & 00185323-4850334 & 4977092322618533760 & K7 \\
4.7238 & -48.0782 & 00185351-4804411 & 4977373140465224320 & M3 \\
4.75 & -53.4972 & 00185991-5329496 & 4924229379808453504 & M3 \\
4.7856 & -50.6626 & 00190849-5039454 & 4972956093019404928 & M2 \\
7.9958 & -65.2795 & 00315893-6516465 & 4708068795401419904 & K7 \\
8.1024 & -62.7352 & 00322472-6244057 & 4900982714699954944 & M2 \\
8.1063 & -63.9459 & 00322550-6356450 & 4900626919608813440 & M1 \\
8.153 & -60.4865 & 00323683-6029093 & 4905178279272755328 & M3 \\
10.3337 & -53.4419 & 00412002-5326301 & 4921838560492998656 & M0 \\
10.3402 & -48.691 & 00412140-4841274 & 4974891680160836224 & M0 \\
10.3834 & -50.6801 & 00413198-5040483 & 4926450668174245248 & M0 \\
10.3967 & -51.5842 & 00413522-5135030 & 4925226980451860480 & M1 \\
10.4117 & -48.3846 & 00413879-4823042 & 4975087599388566912 & M4 \\
10.4179 & -53.9022 & 00414033-5354083 & 4921578594712789888 & M4 \\
10.4233 & -52.2104 & 00414163-5212367 & 4924965468483283200 & M0 \\
15.0549 & -52.1271 & 01001292-5207377 & 4928118111920062848 & M4 \\
15.1272 & -48.6328 & 01003033-4837570 & 4932718296769306112 & M0 \\
15.1354 & -50.7177 & 01003248-5043041 & 4928479198408685568 & M2 \\
15.1636 & -51.2984 & 01003933-5117534 & 4928394845250181120 & M1 \\
15.1732 & -52.6472 & 01004163-5238497 & 4927307256452253312 & M0 \\
15.3229 & -50.3421 & 01011736-5020323 & 4931583944366769408 & M0 \\
15.5803 & -52.7496 & 01021921-5244589 & 4927279047106419584 & M1 \\
15.6506 & -49.0611 & 01023608-4903397 & 4931916993310921600 & K7 \\
15.666 & -49.0829 & 01023983-4904584 & 4931916718433006208 & K7 \\
15.7416 & -51.3862 & 01025798-5123102 & 4928293209144203392 & M2 \\
19.122 & -12.4072 & 01162917-1224252 & 2468686385004236032 & M0 \\
19.1947 & -13.2468 & 01164675-1314482 & 2456488506085006720 & M1 \\
20.069 & -22.5014 & 01201653-2230056 & 5041704225777536256 & M1 \\
20.1538 & -22.3158 & 01203689-2218558 & 5043210934664106240 & M1 \\
20.1576 & -20.6735 & 01203786-2040243 & 2353358679086279936 & M0 \\
22.8801 & -20.9531 & 01313111-2057110 & 5043681525640788096 & K7 \\
22.8953 & -19.5922 & 01313484-1935319 & 5140088732388630912 & M3 \\
22.94 & -19.1524 & 01314551-1909083 & 5140143742332938752 & M3 \\
22.9473 & -20.0364 & 01314735-2002109 & 5043941388341847296 & M1 \\
22.9549 & -18.2411 & 01314916-1814265 & 2450182738180156288 & M0 \\
22.9631 & -19.0234 & 01315107-1901269 & 2449995821203070336 & M2 \\
32.4292 & -19.6991 & 02094293-1941570 & 5137052465388099840 & M2 \\
32.4773 & -23.5008 & 02095450-2330036 & 5121712663273501952 & M4 \\
32.5006 & -20.9393 & 02100005-2056218 & 5124515662009719424 & M1 \\
32.5018 & -23.7051 & 02100040-2342177 & 5121694005935485312 & M1 \\
32.5121 & -23.3846 & 02100287-2323049 & 5123215730028042496 & K7 \\
32.5324 & -23.3426 & 02100777-2320334 & 5123216107985469312 & M0 \\
292.2349 & -40.6332 & 19285636-4037586 & 6737513101292418688 & M0 \\
292.271 & -38.6745 & 19290498-3840268 & 6738700917446630272 & M1 \\
292.2753 & -36.8704 & 19290608-3652135 & 6739515208886313344 & M2 \\
292.2754 & -38.5691 & 19290609-3834083 & 6738703769304980480 & M2 \\
292.3554 & -37.3357 & 19292529-3720085 & 6739211812396448128 & M4 \\
292.3861 & -41.903 & 19293264-4154103 & 6665326692595038976 & M2 \\
292.3878 & -36.6094 & 19293304-3636332 & 6739528196868604416 & M0 \\
292.4241 & -41.2263 & 19294176-4113348 & 6737416275550709504 & M0 \\
292.4722 & -40.9639 & 19295330-4057494 & 6737490939260444544 & M1 \\
292.4735 & -38.846 & 19295360-3850453 & 6737944590885952768 & M1 \\
292.5026 & -38.1511 & 19300061-3809038 & 6739102342271888384 & M2 \\
292.5376 & -39.9457 & 19300904-3956442 & 6737814504918193536 & K7 \\
292.5599 & -41.379 & 19301439-4122435 & 6737400160833990656 & M1 \\
292.5839 & -35.9047 & 19302010-3554168 & 6739630657609761920 & M1 \\
292.5903 & -37.5557 & 19302167-3733204 & 6739193155059715712 & M3 \\
292.6461 & -40.7436 & 19303504-4044356 & 6737500525629064064 & K7 \\
298.3412 & -66.0982 & 19532188-6605530 & 6428334245492474240 & M3 \\
298.3459 & -67.7701 & 19532297-6746125 & 6426888490780555008 & K7 \\
298.3658 & -70.0732 & 19532775-7004235 & 6423365792964956288 & M1 \\
299.8833 & -75.5368 & 19593200-7532130 & 6366722596033434496 & M2 \\
299.8852 & -75.7719 & 19593236-7546189 & 6366506095322532864 & M1 \\
299.9059 & -72.2381 & 19593722-7214164 & 6422137878994211456 & M0 \\
300.7833 & -75.8848 & 20030800-7553053 & 6366511902118845440 & M0 \\
300.8206 & -74.3163 & 20031692-7418577 & 6367678105998927872 & M0 \\
300.8314 & -77.7506 & 20031949-7745017 & 6362732983732674688 & M0 \\
300.8736 & -77.4678 & 20032981-7728039 & 6362933129208830848 & M1 \\
301.3896 & -76.3148 & 20053347-7618532 & 6366440880539147264 & M4 \\
301.3933 & -76.5111 & 20053435-7630399 & 6363409771794337920 & M0 \\
301.4111 & -77.6162 & 20053825-7736585 & 6362928799881750784 & M2 \\
301.4138 & -74.5453 & 20053932-7432433 & 6366908860176040576 & M2 \\
301.5189 & -73.7204 & 20060424-7343130 & 6367734009292578176 & M1 \\
301.6473 & -75.5488 & 20063516-7532543 & 6366535610337681920 & M3 \\
302.1928 & -72.2107 & 20084599-7212373 & 6374109836703866496 & M0 \\
302.1953 & -76.7082 & 20084682-7642291 & 6363391350679444224 & M1 \\
302.2385 & -74.3382 & 20085706-7420164 & 6366936588484228992 & M0 \\
302.3059 & -77.5129 & 20091355-7730461 & 6363116751945430912 & K7 \\
302.4183 & -74.038 & 20094033-7402161 & 6366960502862346880 & M3 \\
302.5245 & -74.3815 & 20100587-7422532 & 6366932911993073536 & M3 \\
304.6379 & -60.5984 & 20183298-6035541 & 6443135286909909888 & M1 \\
304.6475 & -62.7926 & 20183509-6247325 & 6430655177098790656 & K7 \\
304.7004 & -60.5843 & 20184817-6035034 & 6443135596147568128 & M3 \\
304.719 & -60.7144 & 20185249-6042510 & 6443130648345117440 & M1 \\
304.7532 & -63.9155 & 20190090-6354563 & 6429424136396604032 & M1 \\
304.7844 & -62.0634 & 20190823-6203486 & 6430901742581953408 & K7 \\
304.8169 & -64.5867 & 20191605-6435127 & 6429324527512107008 & M0 \\
305.24 & -65.8169 & 20205748-6549005 & 6425930644355539328 & M2 \\
305.2912 & -62.4559 & 20210991-6227212 & 6430689811718803200 & M0 \\
314.3116 & -49.0623 & 20571473-4903441 & 6478282619200079232 & M4 \\
314.3173 & -50.4745 & 20571611-5028284 & 6477958228910103808 & M1 \\
314.3259 & -52.833 & 20571828-5249572 & 6476512577278555008 & M2 \\
314.4073 & -49.6871 & 20573776-4941123 & 6478203042046301952 & M2 \\
314.4262 & -51.3785 & 20574235-5122421 & 6477496262227666304 & M1 \\
314.4476 & -48.2281 & 20574740-4813410 & 6481437289859831936 & M0 \\
314.5364 & -47.9471 & 20580870-4756497 & 6481450380919437568 & M0 \\
314.5708 & -51.7533 & 20581690-5145111 & 6476702346112930560 & M3 \\
314.5861 & -53.3084 & 20582057-5318298 & 6476280236727791488 & K7 \\
314.6217 & -48.2112 & 20582913-4812403 & 6481427119377283072 & M2 \\
316.3029 & -72.238 & 21051269-7214174 & 6371848553602876544 & M4 \\
316.3569 & -73.7044 & 21052563-7342170 & 6370019034972766464 & M0 \\
316.4069 & -74.611 & 21053758-7436391 & 6369827926107617408 & M2 \\
316.5137 & -72.9976 & 21060304-7259507 & 6370253299669186944 & K7 \\
316.5461 & -76.5223 & 21061096-7631203 & 6368358153940497664 & M0 \\
316.6646 & -47.9891 & 21063947-4759203 & 6479978718965033600 & M0 \\
316.6805 & -51.2622 & 21064331-5115441 & 6477067521412743936 & M0 \\
316.6866 & -50.8197 & 21064473-5049102 & 6477188124094148608 & M0 \\
316.7346 & -52.3832 & 21065624-5222585 & 6476760070473571200 & M1 \\
316.7763 & -75.8816 & 21070620-7552536 & 6368951890219260672 & K7 \\
316.7808 & -73.9927 & 21070691-7359330 & 6369952308360570880 & M1 \\
316.8139 & -71.9542 & 21071525-7157151 & 6371881470232372480 & K7 \\
316.8159 & -53.9481 & 21071580-5356527 & 6464234159132043776 & M2 \\
316.8402 & -74.6913 & 21072153-7441291 & 6369823424981789184 & M4 \\
317.0286 & -77.0813 & 21080745-7704525 & 6368255658841466368 & M1 \\
320.8627 & -53.1939 & 21232692-5311380 & 6463767931842885632 & M0 \\
320.8886 & -53.9367 & 21233324-5356122 & 6463663340798479488 & M1 \\
320.8894 & -48.2343 & 21233347-4814037 & 6479185490045244160 & M2 \\
320.8926 & -51.4282 & 21233419-5125414 & 6466216166280290176 & M1 \\
320.9452 & -50.6946 & 21234679-5041394 & 6466320929122308352 & M1 \\
320.9559 & -51.9219 & 21234944-5155189 & 6466094670246298752 & M0 \\
321.0039 & -49.0663 & 21240115-4903577 & 6467116838101821568 & M3 \\
321.2061 & -52.3414 & 21244940-5220288 & 6465331162499653248 & M2 \\
321.2539 & -48.6666 & 21250084-4839587 & 6467152576525084544 & M0 \\
321.2669 & -48.1107 & 21250412-4806365 & 6479179545810551040 & M1 \\
321.2908 & -49.5336 & 21250984-4932015 & 6466952495473971584 & M2 \\
321.3111 & -49.8057 & 21251469-4948205 & 6466930024204871936 & M3 \\
328.0035 & -55.9749 & 21520084-5558294 & 6460607694905782912 & M2 \\
328.0263 & -57.2998 & 21520630-5717594 & 6412368099706810496 & M3 \\
328.0466 & -58.1673 & 21521107-5810017 & 6410788994853191552 & M0 \\
328.1083 & -56.9157 & 21522601-5654563 & 6412415035108541440 & M0 \\
328.1163 & -54.9942 & 21522777-5459378 & 6461064095310355328 & M3 \\
328.1519 & -59.1771 & 21523641-5910368 & 6410433646436042368 & M4 \\
329.507 & -68.1416 & 21580169-6808291 & 6396979575481957120 & M0 \\
329.5357 & -71.6042 & 21580856-7136142 & 6395170569617138432 & M0 \\
329.5857 & -68.3825 & 21582074-6822579 & 6396200915090592128 & M1 \\
329.6402 & -68.532 & 21583342-6831549 & 6396195451892479744 & M0 \\
329.6407 & -66.5214 & 21583359-6631167 & 6398720068028753024 & M0 \\
329.655 & -66.1242 & 21583727-6607269 & 6398734327320523136 & K7 \\
329.6812 & -71.4807 & 21584337-7128496 & 6395172974797882240 & M1 \\
329.6981 & -70.8107 & 21584735-7048373 & 6395349893091578752 & M3 \\
329.7707 & -66.6126 & 21590490-6636446 & 6398705705658038528 & M0 \\
329.7858 & -69.9769 & 21590848-6958366 & 6395961049757161344 & M2 \\
330.1614 & -69.0818 & 22003858-6904533 & 6396119001474821760 & M0 \\
330.2304 & -68.7882 & 22005521-6847171 & 6396176244799223680 & M1 \\
330.4793 & -67.098 & 22015489-6705520 & 6398541053792541312 & M0 \\
330.4881 & -71.1986 & 22015699-7111538 & 6395277050448923520 & M1 \\
330.537 & -66.6793 & 22020880-6640449 & 6398657911261911424 & M1 \\
330.5536 & -69.2018 & 22021296-6912047 & 6396114362910173312 & M0 \\
330.6095 & -71.1017 & 22022638-7106059 & 6395283647515686912 & M0 \\
330.6627 & -69.9695 & 22023900-6958097 & 6395777774913000320 & M2 \\
330.6635 & -69.3764 & 22023917-6922349 & 6396006679490533376 & M1 \\
335.3967 & -49.7882 & 22213517-4947174 & 6511861326356030848 & M1 \\
336.8746 & -65.2209 & 22272993-6513154 & 6404240406674616064 & K7 \\
336.9533 & -61.1955 & 22274878-6111420 & 6406886484485966080 & M0 \\
336.9718 & -62.0438 & 22275317-6202378 & 6406730182035851264 & M1 \\
336.9988 & -65.0326 & 22275990-6501555 & 6404337335496639744 & M2 \\
337.0117 & -63.3252 & 22280284-6319295 & 6405015287494373888 & M0 \\
337.1341 & -63.3473 & 22283209-6320500 & 6405014600299775104 & M3 \\
340.6016 & -55.2521 & 22422438-5515075 & 6506614800465225472 & M1 \\
340.6102 & -58.1029 & 22422639-5806105 & 6503681853198330496 & M0 \\
340.6805 & -56.9029 & 22424330-5654099 & 6504255111073279232 & M0 \\
340.7324 & -55.8439 & 22425580-5550380 & 6505822842855567616 & M0 \\
341.0083 & -55.2843 & 22440202-5517029 & 6505866754601833856 & M3 \\
341.0891 & -57.9258 & 22442136-5755325 & 6504046444382581760 & M2 \\
341.1703 & -57.0863 & 22444083-5705103 & 6504201612960643968 & M0 \\
341.6358 & -59.3226 & 22463257-5919213 & 6503361757875496576 & M0 \\
341.6797 & -58.0314 & 22464302-5801532 & 6503860042801179008 & M1 \\
341.7704 & -57.8642 & 22470480-5751503 & 6503884919251887744 & M0 \\
341.7823 & -59.1236 & 22470768-5907245 & 6503370622688061952 & M1 \\
341.7848 & -57.4271 & 22470830-5725368 & 6503994526817585408 & M0 \\
341.8669 & -57.0766 & 22472806-5704359 & 6504006690164558848 & K7 \\
341.8883 & -57.3651 & 22473318-5721536 & 6503993324226096896 & M0 \\
341.9262 & -56.768 & 22474226-5646044 & 6504227554563443200 & K7 \\
342.058 & -68.1123 & 22481380-6806438 & 6384974901371874432 & M1 \\
342.1365 & -66.5186 & 22483252-6631068 & 6391929003179743104 & M0 \\
342.378 & -70.7325 & 22493063-7043561 & 6384279769504059648 & K7 \\
342.3969 & -71.6922 & 22493560-7141313 & 6383945002574051072 & M1 \\
342.4851 & -67.4008 & 22495623-6724020 & 6391021184532789504 & M2 \\
342.486 & -70.7009 & 22495615-7042016 & 6384278468128958976 & M2 \\
342.5915 & -67.8699 & 22502187-6752114 & 6390987237110516608 & M0 \\
347.3385 & -49.8243 & 23092119-4949276 & 6502996685655270912 & M2 \\
347.3927 & -51.6803 & 23093427-5140489 & 6502361923849262720 & M3 \\
347.4289 & -51.4388 & 23094284-5126190 & 6502394531241135360 & M0 \\
350.4223 & -49.8661 & 23214120-4951580 & 6526074025972604928 & M0 \\
350.4737 & -52.4363 & 23215367-5226106 & 6501389234015189376 & M2 \\
350.4806 & -52.7099 & 23215540-5242347 & 6501290003090808064 & M3 \\
350.5158 & -49.8138 & 23220373-4948491 & 6526074816248262528 & M0 \\
350.5873 & -53.8697 & 23222079-5352097 & 6499715811677467136 & M1 \\
350.6396 & -52.3549 & 23223349-5221174 & 6501395212609739008 & M0 \\
350.702 & -52.7438 & 23224857-5244375 & 6501278492578447488 & M3 \\
350.7246 & -50.56 & 23225391-5033358 & 6501965859144654848 & M3 \\
350.729 & -51.6076 & 23225491-5136273 & 6501811343401430144 & M1 \\
350.9227 & -52.3418 & 23234147-5220307 & 6501381365635183872 & M3 \\
350.9249 & -50.1135 & 23234185-5006485 & 6526019978104413952 & M0 \\
350.9317 & -48.3676 & 23234351-4822035 & 6526632784038094976 & M2 \\
350.9741 & -49.2686 & 23235398-4916062 & 6526162021262815488 & M2 \\
351.0546 & -53.536 & 23241303-5332101 & 6501040757548608256 & M0 \\
351.08 & -50.0411 & 23241898-5002281 & 6526024071208077568 & M4 \\
351.1067 & -51.3368 & 23242553-5120114 & 6501838899911227264 & M0 \\
351.1476 & -48.5501 & 23243541-4832595 & 6526583477813492736 & M3 \\
351.1547 & -53.0622 & 23243699-5303439 & 6501082435911615872 & M3 \\
351.1788 & -49.5513 & 23244272-4933035 & 6526061450308772096 & M3 \\
351.2735 & -48.492 & 23250545-4829313 & 6526584611684870656 & M0 \\
355.4655 & -49.2922 & 23415167-4917321 & 6523162072506258688 & M1 \\
355.4769 & -50.7695 & 23415465-5046097 & 6522600600021061632 & M0 \\
355.5223 & -50.355 & 23420527-5021176 & 6523011164535166720 & M3 \\
358.7286 & -56.0164 & 23545493-5600590 & 6496660371942814464 & M1 \\
358.8308 & -54.0629 & 23551929-5403456 & 6497258987305153920 & M3 \\
358.8433 & -55.2542 & 23552259-5515138 & 6497079938710683904 & M0 \\
358.8753 & -54.7908 & 23553007-5447273 & 6497121479632474240 & M1 \\
358.877 & -54.7158 & 23553034-5442571 & 6497123266338916096 & M0 \\
358.8936 & -54.3689 & 23553433-5422074 & 6497227204546900736 & M2
\label{TabTIC}
\end{longtable}
\bsp
\label{lastpage}
\end{document}